\documentclass[aps,pra,twocolumn]{revtex4-1}

\usepackage{graphicx}
\usepackage{amsfonts}
\usepackage{amsmath}
\usepackage{color}
\usepackage{bm}
\usepackage{amssymb}

\usepackage{multirow}
\usepackage{epstopdf}
\usepackage{natbib}
\usepackage{braket}

\usepackage{url}
\usepackage{soul}

\usepackage{xcolor}
\usepackage{enumitem}

\definecolor{fig_blue}{HTML}{1F77B4}
\definecolor{fig_orange}{HTML}{FF7F0E}
\definecolor{fig_green}{HTML}{2CA02C}

\newcommand{\symLogScale}[1]{Note the symmetric logarithmic scaling of the abscissa, where the linear portion is within ${\pm}#1\tilde t$}
\newcommand{\beq}{\begin{equation}}
\newcommand{\eeq}{\end{equation}}

\def\vec#1{{\bm #1}}

\def\etal{{\it et al.}}

\def\jcp#1#2#3{{\it J.~Chem.~Phys.}~{\bf #1},\ #2\ (#3)}

\def\prl#1#2#3{{\it Phys.~Rev.~Lett.}~{\bf #1},\ #2\ (#3)}

\def\colvecnext#1{
        #1
        \global\advance\colveccount-1
        \ifnum\colveccount>0
                \\
                \expandafter\colvecnext
        \else
                \end{pmatrix}
        \fi
}

\begin{document}

\title{
The effect of anisotropy of long-range hopping on localization in three-dimensional lattices
}
\author{J. T. Cantin, T. Xu, and R. V. Krems}
\affiliation{Department of Chemistry, University of British Columbia, Vancouver, B.C., V6T 1Z1, Canada}

\date{\today}

\begin{abstract}
It has become widely accepted that particles with long-range hopping do not undergo Anderson localization. However, several recent studies demonstrated localization of particles with long-range hopping. In particular, it was recently shown that the effect of long-range hopping in 1D lattices can be mitigated by cooperative shielding, which makes the system behave effectively as one with short-range hopping. Here, we show that cooperative shielding, demonstrated previously for 1D lattices, extends to 3D lattices with \emph{isotropic} long-range $r^{-\alpha}$ hopping, but not to 3D lattices with dipolar-like \emph{anisotropic} long-range hopping. 
We demonstrate the presence of localization in 3D lattices with uniform ($\alpha=0$) \emph{isotropic}   long-range  hopping and the absence of localization with uniform \emph{anisotropic}  long-range hopping by using the scaling behaviour of eigenstate participation ratios. We use the scaling behaviour of participation ratios and energy level statistics to show that the existence of delocalized, non-ergodic extended, or localized states in the presence of disorder depends on both the exponents $\alpha$ and the isotropy of the long-range hopping amplitudes.



\end{abstract}

\maketitle

\section{Introduction}

There is increasing interest in disordered lattice models with long-range hopping, defined as hopping with the amplitude 
$t \propto 1/r^\alpha$, where  $\alpha\leq d$ and $d$ is the dimension of the lattice. The particular case of $\alpha =3$ is exhibited by excitons in molecular crystals with topological
disorder \cite{Logan1987}, organic semiconductors with impurities \cite{Mladenovic2015},
diluted ensembles of atoms and molecules trapped in optical lattices \cite{dy-1,jun-ye-0,jun-ye-1,jun-ye-2,jun-ye-nature,Hazzard2014}, $J$-aggregates \cite{Wuerthner2011}, photo-synthetic complexes \cite{Scholes2011,Fidler2014}, and
ensembles of Rydberg atoms \cite{Robicheaux2014,Muelken2007,Schempp2015}. 

The effect of long-range hopping on Anderson localization was considered already in the original paper of Anderson \cite{anderson}. 
Anderson concluded, via a locator expansion, that particles with long-range hopping do not localize \cite{anderson}. This problem was later revisited by Levitov who examined the resonance behaviour of particles with dipolar hopping in three-dimensional (3D) disordered lattices and concluded that they delocalize because the number of resonances diverges with the lattice size \cite{Levitov1989,Levitov1990,Levitov1990b}.
The arguments of Levitov have recently been applied to study many-body localization of particles with long-range interactions \cite{long-range-effects}.  
It has thus become widely accepted that non-interacting particles with long-range hopping (e.g. $t \propto 1/r$ in 1D lattices or $t \propto 1/r^3$ in 3D lattices) do not undergo Anderson localization. 

The generality of this conclusion has, however, been questioned by several recent authors. For example, Deng \etal~\cite{Deng2018} have used exact diagonalization and analysis of the multi-fractal spectrum to demonstrate algebraic localization for a particle with long-range hopping in a 1D system with off-diagonal disorder. Nandkishore and Sondhi have shown using field-theory techniques that many-body systems with long-range interactions in 1D and 2D can localize  and hypothesized that the same is true in 3D \cite{Nandkishore2017}. Nandkishore and Sondhi even go so far as to question the validity of the locator expansion and resonance arguments for systems with long-range hopping.


Furthermore, Santos \etal~\cite{Santos2016} and Celardo \etal~\cite{Celardo2016} have discovered a phenomenon, cooperative shielding, that causes effective short-range behaviour in a system with long-range interactions or hopping. Cooperative shielding precludes the resonance arguments used to demonstrate delocalization in the presence of on-site disorder and allows localization to take place. Celardo \etal~\cite{Celardo2016} have in particular demonstrated localization in 1D for particles with long-range hopping. Ossipov has also demonstrated localization of particles with  uniform hopping on a \emph{d}-dimensional simplex, via a similar mechanism \cite{Ossipov2013}.

In an earlier paper, Burin and Maksimov \cite{Burin1989} predicted the localization of particles with long-range hopping via a renormalization procedure. The authors suggested that their conclusion was different from that of Levitov \cite{Levitov1989} because of the different symmetries of the hopping amplitudes considered: Levitov explored the system with \emph{anisotropic} dipolar hopping, while Burin and Maksimov considered \emph{isotropic} long-range hopping. All of the more recent papers demonstrating the localization of particles with long-range hopping have considered isotropic hopping.
However, the effect of anisotropy of long-range hopping in high-dimensional lattices has not been examined. 

Here, we show that the cooperative shielding demonstrated previously for 1D lattices extends to particles with \emph{isotropic} long-range hopping in 3D systems. However, this cooperative shielding does not extend to particles in 3D systems with \emph{anisotropic} dipolar hopping. 
We also provide numerical and analytical evidence for the localization of particles with long-range isotropic hopping in 3D lattices. 

The remainder of the paper is organized as follows: Following the description of the models, 
we discuss the phenomenon of cooperative shielding in 3D systems. We analytically diagonalize the Hamiltonian with isotropic hopping $t \propto 1/r^\alpha$ for 3D lattices for arbitrary $\alpha$ and ilustrate that it exhibits the same energy level structure as that of a 1D system with cooperative shielding. We find this to be the case for both periodic and open boundary conditions. The presence of cooperative shielding suggests the presence of localization. Contrastingly, we find no clear evidence for cooperative shielding at most values of $\alpha$ for anisotropic dipolar hopping, for either periodic or open boundary conditions. 

We use the scaling behaviour of participation ratios to demonstrate the existence of localized states for  isotropic hopping when $\alpha = 0$ (i.e.~uniform infinite-range hopping), even for weak disorder. We show that the anisotropic hopping case contains only non-ergodic extended and delocalized states, even for strong disorder. This provides further evidence that cooperative shielding does not exist for the anisotropic hopping considered here, as cooperative shielding is expected to be strongest when $\alpha = 0$. 

We then examine finite $\alpha$, and show that anisotropic hopping again only supports delocalized and non-ergodic extended states for $\alpha =1$. Interestingly, for anisotropic hopping with $\alpha=3$ (i.e. dipolar interactions), we observe no delocalized states given sufficient disorder. Whether the states are localized or non-ergodic extended is indeterminate, however. The isotropic hopping case supports delocalized states and either non-ergodic extended or localized states (or both) for $\alpha =1 \text{ and } 3$. 

Following this, we use the scaling behaviour of energy level statistics to investigate the physically important case of $\alpha =3$. Our results indicate the presence of at least some localized states near zero energy for isotropic hopping, while the anisotropic case is inconclusive. To connect with currently attainable experiments, we also examine the phase diagram for both diagonal and binary off-diagonal disorder (as per the quantum percolation model). This provides a basic map that can guide experiments with relevant finite systems, such as polar molecules on an optical lattice \cite{jun-ye-nature}.




\section{Models}
We consider a single particle in a disordered and diluted cubic lattice with $N$ sites per dimension, with the hopping amplitude
 $ \propto \frac{1}{r^\alpha}$, where $\alpha$ sets the hopping range. 
The angular dependence of the hopping amplitude is described below. 
The Hamiltonian we consider has the following general structure:
\begin{eqnarray}
\hat H = \sum_{i} \omega_i \hat c^\dagger_i \hat c_i +  \sum_{i} \sum_{j \neq i} t_{ij} \hat c^\dagger_i \hat c_j, 
\label{model}
\end{eqnarray}
where the operator $\hat c_i$ removes the particle from site $i$, $\omega_i$ is the energy of the particle in site $i$, and $t_{ij}$ is the amplitude for particle tunnelling from site $j$ to site $i$. We introduce disorder by randomizing both the values of $\omega_i$ and $t_{ij}$, which makes Eq.~(\ref{model}) relevant for both disordered lattices and amorphous systems.

We randomize the values $\omega_i \in [-\omega/2, \omega/2]$ by drawing them from a uniform distribution. We define the disorder strength as 
\begin{eqnarray}
W \equiv \frac{\omega}{t_{\text{max}}},
\end{eqnarray}
 where $t_{\text{max}} \equiv \max |t_{ij}|$, in order to allow a direct comparison between the isotropic and anisotropic models defined below by normalizing the disorder amplitude to the largest hopping amplitude present. 

We randomize the values $t_{ij}$ as in the site percolation model. To do this, we define  $t_{ij} = d_{ij}\tau_{ij}$, where $d_{ij}$ is the dilution parameter, and divide the lattice sites into two subsets $P$ and $Q$, with $p N^3$ sites in the $P$ subset and $(1-p)N^3$ in the $Q$ subset. For a given value $p$, the lattice sites are assigned to the subsets at random. The dilution parameter is then defined as:
\begin{align}
d_{ij} &= \begin{cases} 
1, & i \in P \text{ and } j  \in P \\
0, & i \in Q \text{ and/or } j  \in Q
\end{cases} \label{dijDef}
\end{align}

With $t_{ij}$ and $d_{ij}$ thus defined, the Hamiltonian (\ref{model}) describes a generic particle in a disordered, diluted lattice with $pN^3$ sites. 
The angular dependence of the hopping amplitude is determined by the magnitudes of $\tau_{ij}$. We consider two types of models for $\tau_{ij}$. 

\subsection{Isotropic Hopping}

For the case of isotropic hopping, we define 
\begin{eqnarray}
\tau^\text{I}_{ij} =\frac{\gamma}{|\vec{r}_{ij}|^{\alpha}},
\label{model_iso}
\end{eqnarray}
where $\bm{r}_{ij} = \bm{r}_{i} - \bm{r}_j$ is the distance between sites $i$ and $j$ in the 3D lattice. With $\alpha > 3$ the hopping is short-range, with $\alpha \leq 3$ -- long-range. The value of $\gamma$ is chosen such that $\tau^\text{I}_{ij} \equiv \tilde t = 1$ for nearest neighbour (NN) sites. Thus, $t_{\text{max}}$ is 1 for isotropic hopping. 


\subsection{Anisotropic Hopping}

We choose the tensorial form of the anisotropic hopping to be the same as the dipole - dipole interaction between polar molecules subjected to an electric field along the $z$-direction, see, for example, Ref. \cite{tianrui}: 
\begin{align}\tau^\text{A}_{ij} =\frac{\gamma\left(1-3\cos ^2 \theta_{ij}\right)}{|\vec{r}_{ij}|^{\alpha}},
\label{model_aniso}
\end{align}
where $\theta_{ij}$ is the angle between $\vec{r}_{ij}$ and the $z$-axis. We choose $\gamma$ as defined above, which makes $\tau^\text{A}_{ij} = 1$ for NN sites in the $x$-$y$ plane and $\tau^\text{A}_{ij} = -2$ for NN sites along the $z$-axis. Thus, $t_{\text{max}} = 2$ for the anisotopic hopping considered here. 


In this article, the system parameters of interest are $p$, $W$, $N$, $\alpha$ and whether the hopping is iso- or anisotropic. For all of the numerical results, we perform exact diagonalization to obtain \emph{all} of the eigenvalues and eigenstates of the dense Hamiltonian matrix. These eigenvalues and eigenstates are needed to perform the energy level statistics without any approximations and to examine the participation ratios across the entire spectrum. This restricts the size of the system to be smaller than what can be achieved with state-of-the-art Jacobi-Davidson algorithms which only obtain a small subset of the total number of eigenvalues and eigenstates \cite{Vasquez2008,Rodriguez2008}. However, we are still able to observe scaling behaviour for some properties.

\section{Cooperative Shielding in 3D}
Recent work has shown the presence of cooperative shielding in 1D single- and many-body systems with long-range hopping or interactions \cite{Celardo2016,Santos2016}. Cooperative shielding allows the dynamics of a Hamiltonian with long-range features to be described by an effective short-range Hamiltonian for a finite time, i.e. the dynamics are effectively ``shielded'' from the long-range components for a finite time. 

This phenomenon occurs because of the formation of short- and long-range subspaces in the Hilbert space, in turn caused by the long-range terms in the Hamiltonian. In many instances, the short-range subspaces have many more states than, and are separated by an energy gap from, the long-range subspace. When this gap increases with the system size, it increases the ``shielding'' time and makes the shielding cooperative. In the infinite size limit, the dynamics of a wavefunction initially in the short-range subspace are then completely governed by the effective short-range Hamiltonian. Cooperative shielding allows for the  localization of particles with long-range hopping \cite{Celardo2016} and provides an explanation as to why localization is observed in various systems with long-range hopping or interactions \cite{Deng2018,Nandkishore2017,Burin1989}.

Here, we show that the energy level structure and gap behaviour conducive to cooperative shielding in 1D is also present in 3D, if the long-range hopping is \emph{isotropic}. If the hopping is \emph{anisotropic}, the energy level structure becomes different and does not exhibit well-separated Hilbert subspaces, precluding the cooperative shielding described in \cite{Celardo2016}.

\begin{figure*}[ht]
	\begin{center}
		\includegraphics[scale=0.4]{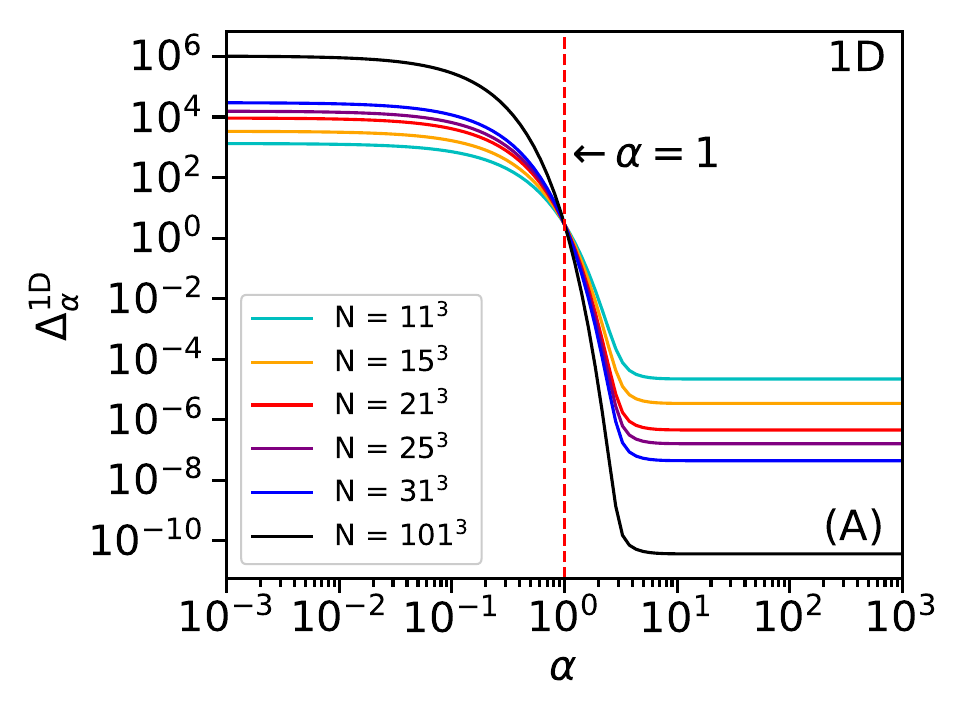}
		\includegraphics[scale=0.4]{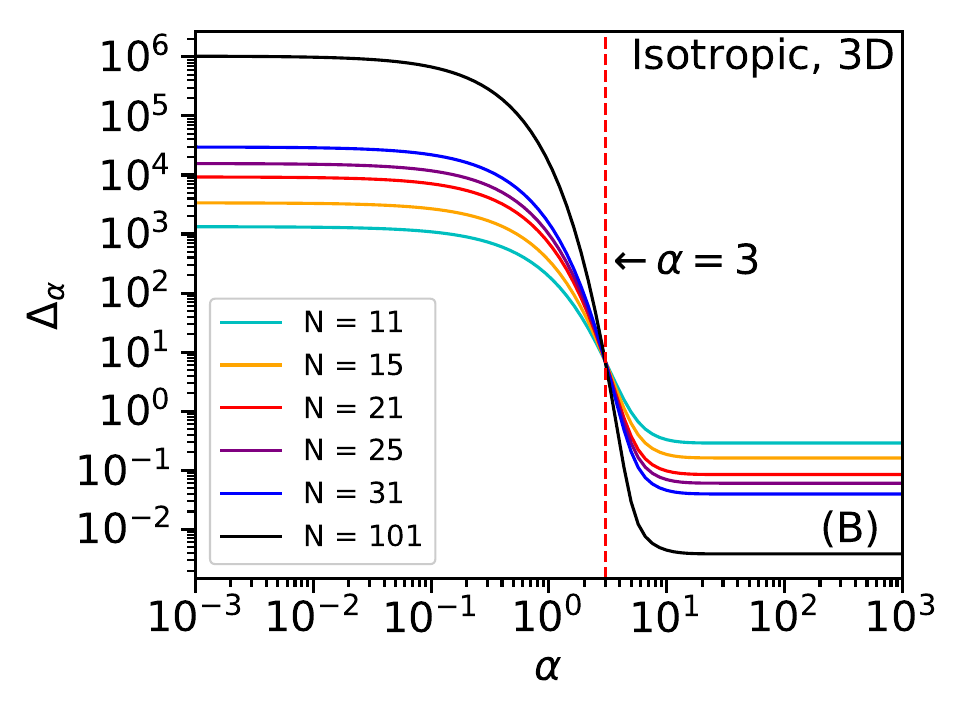} \\
		\includegraphics[scale=0.4]{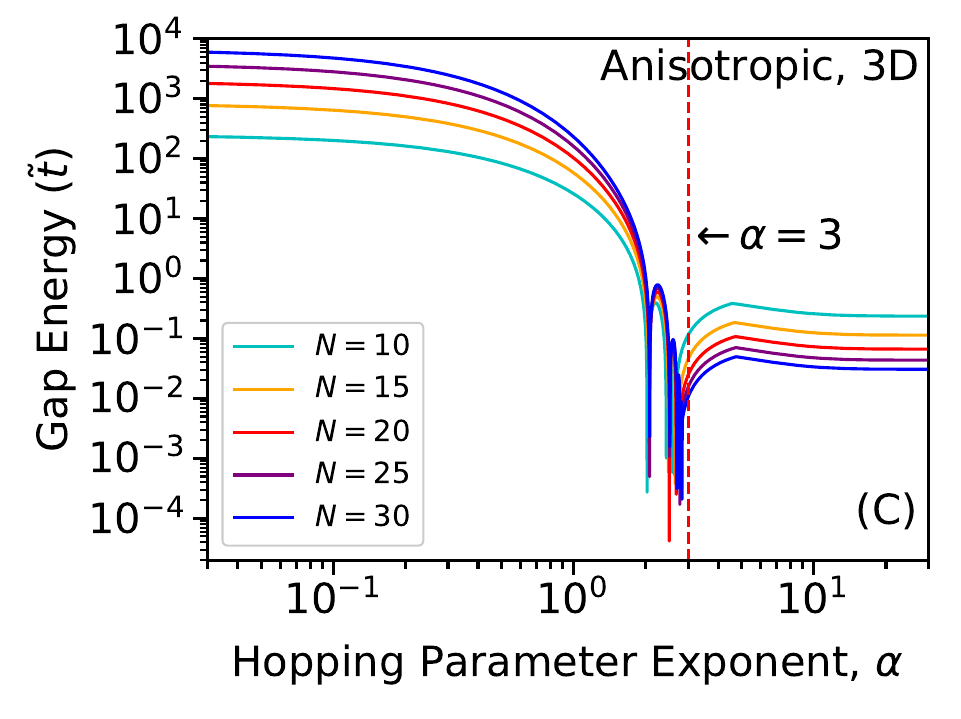} 
		\includegraphics[scale=0.4]{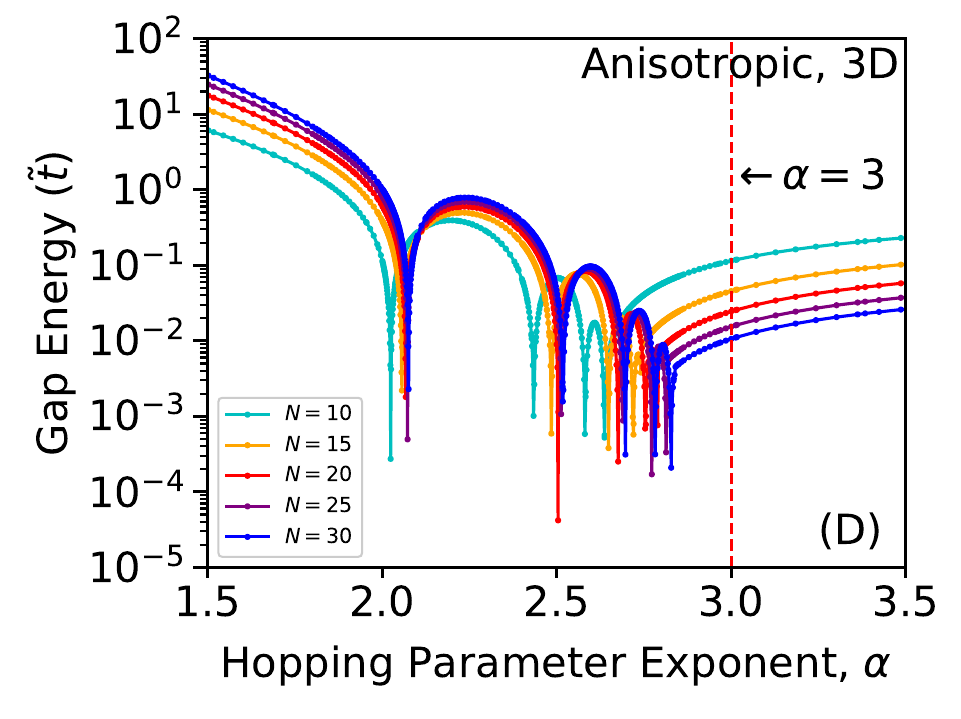} 
	\end{center}
	\caption{ The energy gap at the top of the spectrum as a function of the hopping range exponent $\alpha$. Panel A is for long-range hopping in a 1D lattice with periodic boundary conditions. Panel B is for a 3D lattice with isotropic hopping,  determined analytically under periodic boundary conditions; see Eqn.~(\ref{deltaDef}). Panels C and D show the energy gap for the Hamiltonian with anisotropic hopping, determined numerically and under open boundary conditions. Panel D is an expanded view of panel C. $N$ is the system side-length (total of $N^3$ sites in the 3D case). In all plots, $p=1$ and $W=0$. Note that we examine the top of the spectrum as the effective mass is negative.
	}
	\label{gapPlots}
\end{figure*}

\subsection{Isotropic Hopping} \label{iso_analytics}

To illustrate cooperative shielding in 3D, we first diagonalize analytically the Hamiltonian for an ideal lattice with hopping of arbitrary range. 
The model with the isotropic hopping (\ref{model_iso}) can be written, in the absence of disorder and with periodic boundary conditions, as
\begin{align}
\hat H^I_\alpha &=  \sum_{i,j,h,l,m,n = -\frac{N-1}{2}}^{\frac{N-1}{2}} \frac{\zeta}{r^\alpha}\left(1-\delta_{il}\delta_{jm}\delta_{hn}\right) \hat c_{lmn}^{\dagger} \hat c_{ijh},
\end{align}
where $\zeta > 0$ is the hopping parameter,  the lattice side-length $N$ is assumed odd for simplicity, $\hat c_{ijh}$ is the annihilation operator on site $(i,j,h)$, $r = \sqrt{\left(l-i\right)^2+\left(m-j\right)^2+\left(n-h\right)^2}$ is the unit-less distance between sites, and the term $\frac{\left(1-\delta_{il}\delta_{jm}\delta_{hn}\right)}{r^\alpha}$ is understood to be zero when $i=l$, $j=m$, and $h=n$ (i.e. the particle can hop anywhere except to its original site). Note that $\zeta > 0$ implies a negative effective mass, such as is seen for holes in a semi-conductor. Given that we examine only non-interacting particles, whether the effective mass is positive or negative does not change the final results, except for where one would expect to find the relevant shielding gap (i.e.~at the top or bottom of the spectrum).

The case of $\alpha = 0$ produces the uniform, isotropic Hamiltonian $\hat H^I_0$, which can be diagonalized in the momentum basis to yield
\begin{align}
\hat H^I_0 &= \zeta N^3 \hat c_{000_k}^{\dagger} \hat c_{000_k} - \zeta \sum_{k_1k_2k_3}\hat c_{k_1k_2k_3}^{\dagger} \hat c_{k_1k_2k_3}, 
\label{HuniformDiagd}
\end{align}
where the subscript of $\hat c_{000_k}$ means  $k_1=k_2=k_3=0$. The energy gap between this specific state $\ket{000_k}$ and the $(N^3 -1)$-fold degenerate states $\ket{k_1k_2k_3}$, with one or more of  $k_1,k_2,k_3$ not equal to zero, is thus $\Delta = \zeta N^3$ for the uniform, isotropic Hamiltonian. This is the energy gap responsible for cooperative shielding. In the 1D case, this gap is $\Delta = \zeta N$ \cite{Celardo2016}, which allows us to surmise that in the 2D case, $\Delta = \zeta N^2$ is likely true. The above energy level structure (\ref{HuniformDiagd}) has also been seen by Ossipov for particles with  uniform isotropic hopping on the \emph{d}-dimensional simplex \cite{Ossipov2013}.

The limiting case of $\alpha \to \infty$ produces the tight-binding model, also diagonal in momentum space. Given that the eigenstates of these two limits are momentum states and that the $\frac{1}{r^\alpha}$ factor does not break any additional symmetries, the eigenstates of $\hat H^I_\alpha$ must also be momentum states.

\begin{widetext}
Evaluating the matrix elements of  $\hat H^I_\alpha$ in momentum space, we find
\begin{align}
&E_{k_1,k_2,k_3}^I(\alpha)  = \braket{k_1k_2k_3|\hat H^{I}_{\alpha}|k_1k_2k_3} = \nonumber \\ 
&2 \zeta \sum _{l=1}^{N-1} \left(1-\frac{l}{N}\right) \frac{\cos \left(\frac{2 \pi  k_1 l}{N}\right)+\cos \left(\frac{2 \pi  k_2 l}{N}\right)+\cos \left(\frac{2 \pi  k_3 l}{N}\right)}{l^{\alpha }}  \nonumber \\
&+4 \zeta \sum _{l=1}^{N-1} \sum _{m=1}^{N-1}\left(1-\frac{l}{N}\right) \left(1-\frac{m}{N}\right) \frac{\cos \left(\frac{2 \pi  k_1 l}{N}\right) \cos \left(\frac{2 \pi  k_2 m}{N}\right)\!+\!\cos \left(\frac{2 \pi  k_2 l}{N}\right) \cos \left(\frac{2 \pi  k_3 m}{N}\right)\!+\!\cos \left(\frac{2 \pi  k_3 l}{N}\right) \cos \left(\frac{2 \pi  k_1 m}{N}\right)}{\left(l^2+m^2\right)^{\alpha /2}} \nonumber \\
&+8 \zeta \sum _{l=1}^{N-1} \sum _{m=1}^{N-1} \sum _{n=1}^{N-1} \left(1-\frac{l}{N}\right) \left(1-\frac{m}{N}\right) \left(1-\frac{n}{N}\right)\frac{ \cos \left(\frac{2 \pi  k_1 l}{N}\right) \cos \left(\frac{2 \pi  k_2 m}{N}\right) \cos \left(\frac{2 \pi  k_3 n}{N}\right)}{\left(l^2+m^2+n^2\right)^{\alpha /2}},
\label{Energy_iso}
\end{align}
where $k_i \in \left \{ k | k\in\mathbb{Z} \wedge k\in[-\frac{N-1}{2},\frac{N-1}{2}]\right \}$ refers to the component of the reciprocal lattice vector in the direction $i$, such that the crystal momentum is $2\pi k_i/Na$, $a$ is the lattice constant, and $\ket{k_1k_2k_3}$ is a momentum eigenstate.
\end{widetext}
It can be shown that Eqn.~(\ref{Energy_iso}) reduces to $\braket{k_1k_2k_3|\hat H^{I}_{0}|k_1k_2k_3}$, see Eqn.~(\ref{HuniformDiagd}), in the limit where $\alpha \to 0$ and to the tight-binding model result in the limit where $\alpha \to \infty$ and $N \to \infty$.

The gap for generic $\alpha$ can be defined as 
\begin{eqnarray}
\Delta_\alpha = E_{0,0,0}^I(\alpha) - E_{0,0,1}^I(\alpha)
\label{deltaDef}
\end{eqnarray} 
Given Eqn.~(\ref{Energy_iso}) and $\zeta > 0$, it is clear that $E_{0,0,0}^I(\alpha)$ is the largest eigenvalue and that the next highest energy level has states with a single unit of momentum. This energy level is six-fold degenerate $(\text{i.e.}~ E_{0,0,\pm1}^I=E_{0,\pm1,0}^I=E_{\pm1,0,0}^I, \forall \alpha)$, as can be seen from the symmetries of $E_{k_1,k_2,k_3}^I(\alpha)$ in $k_i$; the choice of $E_{0,0,1}^I(\alpha)$ in the definition of $\Delta_\alpha$ is arbitrary.   Also, since $\zeta > 0$, $E_{0,0,0}^I \geq E_{0,0,1}^I$, and $\Delta_\alpha \geq 0$.

For comparison, in 1D the Hamiltonian is given by: 
\begin{align}
\hat H^{I,\text{1D}}_{\alpha} = \sum_{i,j = -\frac{N-1}{2}}^{\frac{N-1}{2}} \frac{\zeta}{|i-j|^\alpha}\left(1-\delta_{ij}\right) \hat c_{i}^{\dagger} \hat c_{j}.
\label{Hisotropic_1D}
\end{align}
The corresponding eigenvalues are:
\begin{align}
&E_{k}^{I,\text{1D}}(\alpha)  = \braket{k|\hat H^{I,\text{1D}}_{\alpha}|k} = 
2 \zeta \sum _{l=1}^{N-1} \left(1-\frac{l}{N}\right) \frac{\cos \left(\frac{2 \pi  k l}{N}\right)}{l^{\alpha }}, 
\label{Energy_iso_1D}
\end{align}
where $k \in \left \{ k | k\in\mathbb{Z} \wedge k\in[-\frac{N-1}{2},\frac{N-1}{2}]\right \}$ is the component of the reciprocal lattice vector. A corresponding gap can be defined as $\Delta_\alpha^{\text{1D}} = E_{0}^{I,\text{1D}}(\alpha) - E_{1}^{I,\text{1D}}(\alpha)$ and is shown in Figure \ref{gapPlots} (A). This 1D Hamiltonian (\ref{Hisotropic_1D}) has been shown to exhibit cooperative shielding in Ref. \cite{Celardo2016}.

Figure \ref{gapPlots} shows $\Delta_\alpha$ as a function of $\alpha$ for various system sizes.  Panel (A) of Figure \ref{gapPlots} plots the results for the 1D lattice. Panel (B) shows that, for a 3D lattice with isotropic hopping, the dependence of the gap separating the long-range and short-range subspaces on $\alpha$ is qualitatively the same as in 1D, except that the gap becomes independent of the system size at $\alpha = 3$, instead of $\alpha = 1$. We have confirmed by numerical calculations that this behaviour is the same for even and odd $N$ and with open boundary conditions. Based on the 1D and 3D results of Figure 1, it should be expected that a 2D system with isotropic hopping likely has a similar gap behaviour, exhibiting the transition at $\alpha = 2$.



The Anderson transition is known to exist in 3D for $\alpha > 3$ \cite{anderson}, but questions arise for the case with $\alpha \leq 3$. Figure \ref{gapPlots} (B) shows that cooperative shielding can be expected for {\it isotropic} hopping with $\alpha < 3$, suggesting that the Anderson transition could also occur in this region. As a side note, the localization may not be exponential, as seen in \cite{Deng2018}. 

Note that the curves in Figure 1 (B) corresponding to different lattice sizes all converge at the transition point. Thus, 
as in 1D \cite{Celardo2016}, the gap is independent of lattice size when $\alpha$ equals the dimension (in our case, three).
If the disorder strength is smaller than the gap, then the gap between the long- and short-range subspaces remains open. 
A finite gap would seemingly indicate a finite shielding time, perhaps precluding localization at infinite times for $\alpha = 3$. However, consider a quantum particle placed at an individual lattice site. The contribution of the delocalized state $\ket{000_k}$ to the wave packet of this particle is proportional to $\frac{1}{N}$, indicating a zero contribution in the infinite lattice size limit. This leaves the possibility of cooperative shielding and localization open in the infinite size limit, but does not confirm them. In the following sections, we examine the effect of cooperative shielding on localization in 3D lattices. 


\begin{figure}[ht]
	\begin{center}
		\includegraphics[scale=0.4]{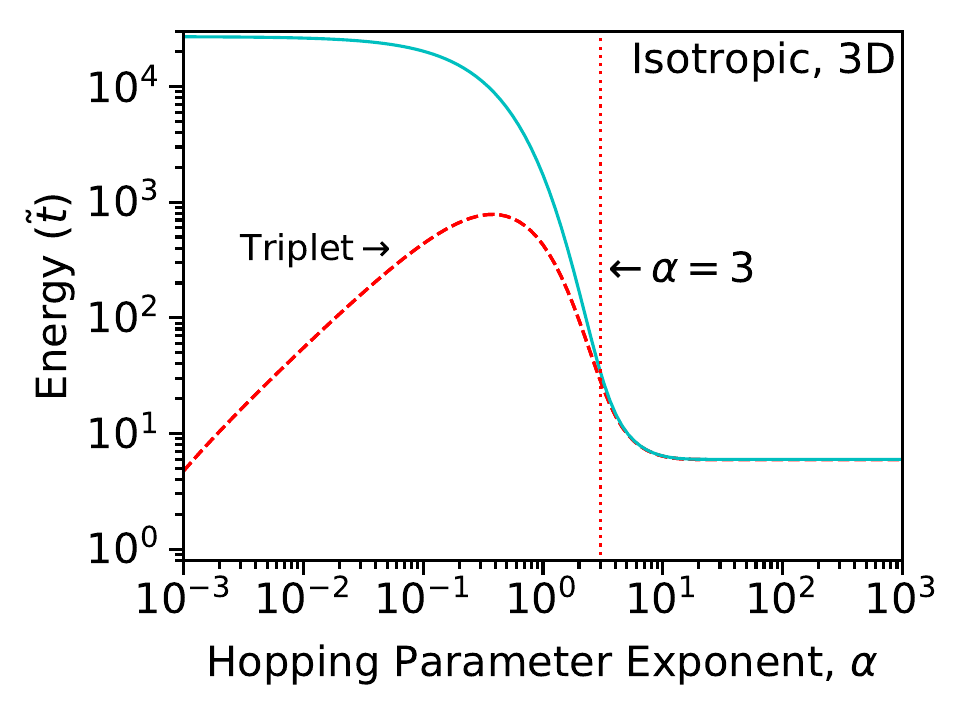} \\
		\includegraphics[scale=0.4]{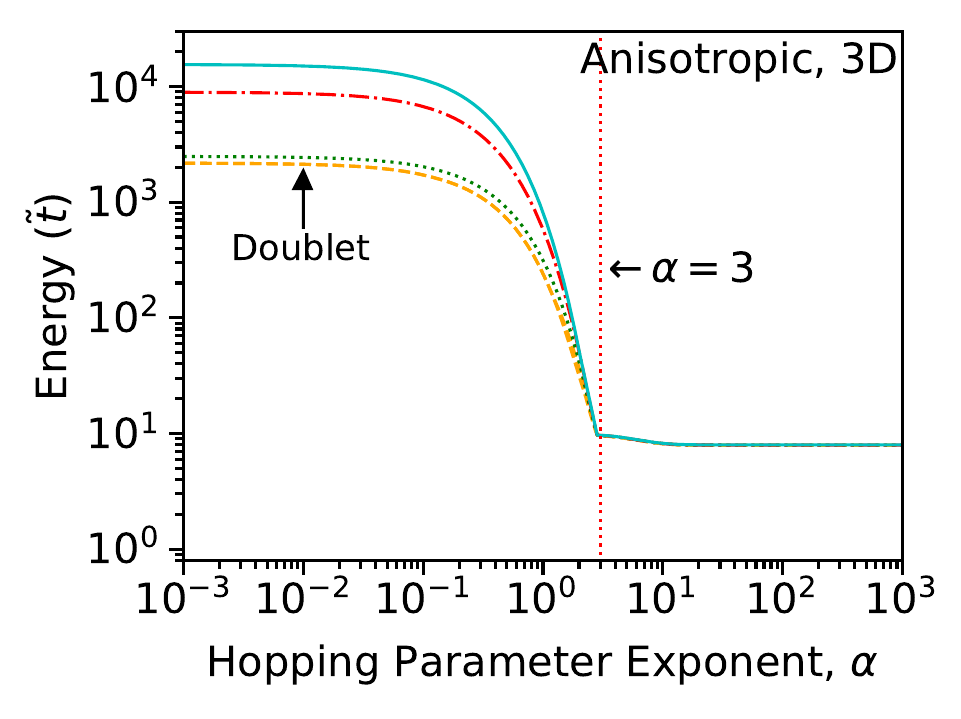} \\
	\end{center}
	\caption{ Energy of several top eigenstates as a function of $\alpha$ for isotropic hopping and anisotropic hopping  with open boundary conditions (top and bottom panels, respectively). For the top panel, the red dashed line is triply degenerate; the other is non-degenerate. For the bottom panel, the orange line is doubly degenerate for $\alpha \lesssim 1$; the rest are non-degenerate. In both plots, $N^3 = 30^3$, $p=1$, and $W=0$.
	}
	\label{energyLevelStructure}
\end{figure}

\subsection{Anisotropic Hopping}

The dependence of the gap separating the short- and long-range subspace is markedly different for the model with anisotropic hopping (see Figure 1 (C) and (D)). 
It is more difficult to obtain the analytical results for the model with anisotropic hopping because of
the $\cos^2 \theta$ term, even if the eigenstates of the system with periodic boundary conditions are still momentum states (rotational symmetry is broken, but not translational).
Here, we study the behaviour of the gap at the top of the spectrum as a function of $\alpha$ numerically, using  open boundary conditions. The results shown in Figure \ref{gapPlots} (C) illustrate the qualitative difference from the isotropic case. In particular, for the anisotropic model, 
there are several values of $\alpha < 3$ where the gap becomes zero. At these values of $\alpha$, cooperative shielding cannot occur so there is no possibility of localization. 

From examining the energy level structure, it can be concluded that the zero-gap points arise from level crossings. It is thus clear that there is no separation of subspaces at the top of the spectrum, in contrast to the case of isotropic hopping. 
While there is a gap that grows with the system size for $\alpha \lesssim 2$, the state at the top of the spectrum is not an equal superposition of all sites, as it is when cooperative shielding is known to occur. Furthermore, the top energy level is shown to be a doublet under periodic boundary conditions, indicating that this particular gap actually disappears in the infinite size limit when translational invariance is restored.

There is, however, a gap between the highest-energy doublet and the next doublet. We examine this gap for $N=11$ and periodic boundary conditions. 
The curve (not shown) appears similar to those in Figure \ref{gapPlots} (C) and (D), though the number of zero-gap points is reduced. This is likely because the higher symmetry permits the formation of doublets, which produce fewer level crossings (i.e.~one for each doublet instead of one for each state). Given these level crossings, the same conclusions can be reached for the periodic boundary case as for the open boundary case: there is no clear separation of subspaces at the top of the spectrum and no evidence for cooperative shielding. While there is a gap at small values of $\alpha$, the states at the top of the spectrum are not equal superpositions of all sites and not necessarily fundamentally different from the rest of the eigenstates. Cooperative shielding, of the form discussed in \cite{Celardo2016}, thus appears precluded. This is investigated more fully further below by examining the scaling behaviour of the participation ratios.

In addition to the gap behaviour, the overall energy level structure appears different between the two cases, as exemplified in Figure \ref{energyLevelStructure} for ideal lattices of side-length $N =30$. Here, the top several eigenstates are shown for both isotropic (top panel) and anisotropic (bottom panel) hopping with open boundary conditions. The values are obtained numerically. Both the degeneracies of the levels and the overall behaviour as a function of $\alpha$ are very different. The energy level crossings in the anisotropic hopping case are in the region $2 \lesssim \alpha \lesssim 3$. Their exact locations are best observed as the zero-gap points in Figure \ref{gapPlots} (D). Note that the differences between Figure \ref{energyLevelStructure} (top panel) and Eqn. (\ref{Energy_iso}), particularly in terms of the degeneracy layout, are due to the differing boundary conditions (open vs. periodic).



 \begin{figure}[ht]
	\begin{center}
		{\bf\large{~~~~~Isotropic, $\bm{\alpha = 0$}}}\\
		\includegraphics[scale=0.4]{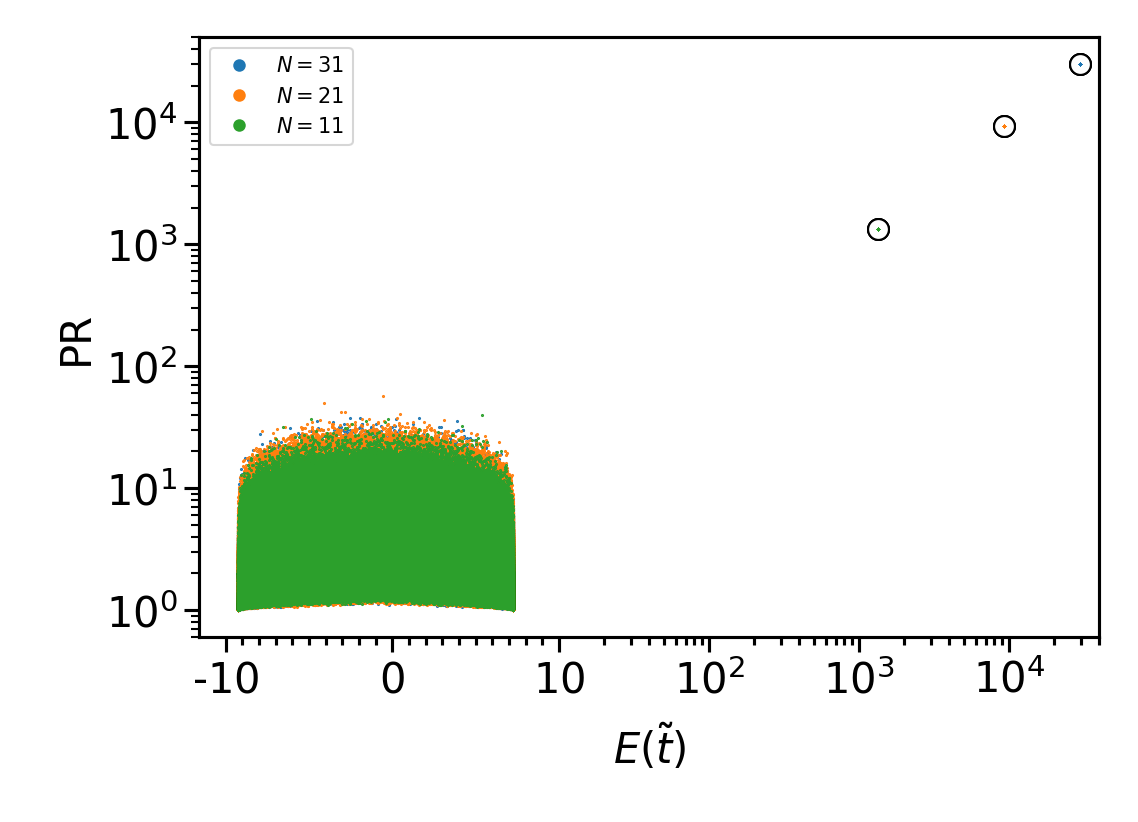} \\
		\includegraphics[scale=0.4]{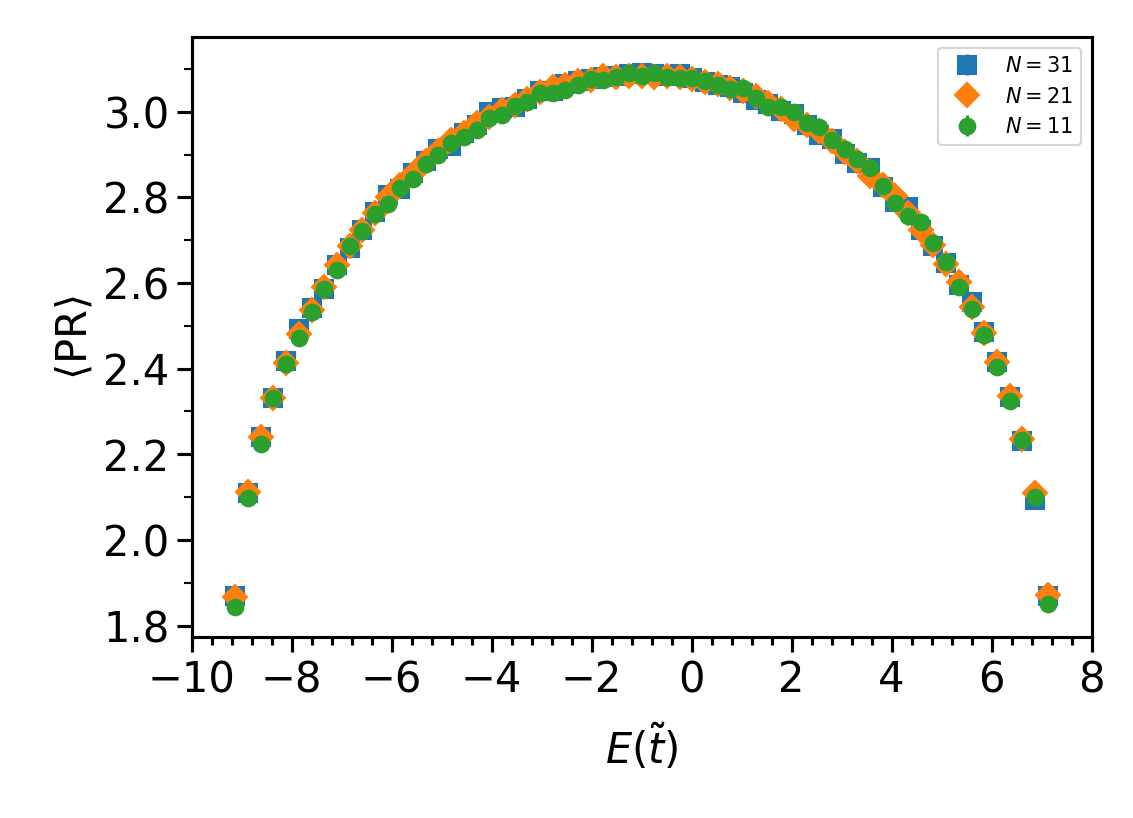} \\
	\end{center}
	\caption{Participation ratio (PR) as a function of Energy for the eigenstate of a 3D system with isotropic hopping, $W = 16.5$, $p=1$, and $\alpha=0$. The number of disorders included ranges from 148 to 2100, as required to obtain sufficiently small error bars. \emph{Top Panel}: Plot of the PR of each eigenstate vs its eigenvalue for several disorder realizations for $N = 11,~21,\text{ and } 31$ (green \textcolor{fig_green}{$\blacksquare$}, orange \textcolor{fig_orange}{$\blacksquare$}, and blue \textcolor{fig_blue}{$\blacksquare$}, respectively). \symLogScale{10}. Large black circles merely highlight the location of the states at the top of the band. \emph{Lower Panel}: PR averaged within energy bins and over disorder realizations. The plot is focused on the band near zero energy. Note the linear scale of the abscissa. Error bars for the average PR are 95\% confidence intervals and smaller than the marker size where not seen.
	}
	\label{PRalphaZeroIsotropic}
\end{figure}

\section{Participation Ratios}

We have established the presence of cooperative shielding in 3D systems with isotropic long-range hopping, which suggests the possibility of localization. We now examine this possibility, for both isotropic and anisotropic hopping with open boundary conditions and $\alpha =0,1,\text{ and }3$, from a participation ratio point of view. In a later section we examine the energy level statistics for $\alpha =3$.

The participation ratio (PR) is an effective count of the number of sites occupied by a wavefunction. It is defined as
\begin{equation}
\text{PR} = \frac{1}{\sum_i |\psi(\vec x_i)|^4},
\label{PRdef}
\end{equation}
where the sum is over all lattice sites, $\psi$ is a normalized wavefunction and $\vec x_i$ is the position of lattice site $i$.

Typically, $\text{PR} \propto N^\beta$ \cite{LevyFlights,DeLuca2014,Mirlin2000}. This scaling behaviour gives information on the nature of the eigenstates. In particular, if $\beta = 0$, the eigenstates are localized, if $\beta = 3$, delocalized and, if $0<\beta<3$, the states can be labelled as extended non-ergodic \cite{LevyFlights,DeLuca2014,Mirlin2000}. Formally, extended non-ergodic states are states where $\text{PR}^{-1}\not\to\langle\text{PR}^{-1}\rangle$ as $N\to\infty$, where $\langle\cdots\rangle$ indicates the average over disorder realizations. Informally and in the context of this paper, extended non-ergodic states span the system, but do not fill the entirety of the accessible space. Extended non-ergodic states are typically multi-fractal, though this property is definitively identified by examining the generalized participation ratios \cite{DeLuca2014,Mirlin2000,Janssen1998}. We focus on the scaling behaviour of the PR (\ref{PRdef}), rather than the generalized participation ratios, as the limited system sizes accessible when obtaining the entire specturm via exact diagonalization makes a full quantitative analysis difficult.

We analyze the scaling behaviour of the PR of the eigenstates via two approaches. The first obtains a quantitative estimate of $\beta(E)$ by fitting the above scaling equation to the behaviour of $\langle\text{PR}\rangle$, where $\langle\text{PR}\rangle$ is the PR averaged within a particular energy window and over multiple disorder realizations. The second approach obtains a qualitative understanding of the system by examining the scaling behaviour of the distributions of both PR and PR$/N^3$ at all energies. Three cases are discerned: 
\begin{enumerate}[label=(\roman*)]
\item If the PR \emph{does not} change with the system size, but PR$/N^3$ \emph{does}, then $\beta = 0$.
\item If the PR \emph{scales} with the system size, but PR$/N^3$ \emph{does not}, then $\beta = 3$.
\item If \emph{both} PR and PR$/N^3$ change with the system size, then $0<\beta<3$. 
\end{enumerate}
No other cases arise as $\beta \in [0,3]$.


We note that the spectra of the systems we examine have an interesting dual-scale behaviour: there is a large density of states near zero energy, but also many states at energies very far away from zero. Thus, the spectra typically have one or more ``wings''. To capture the behaviour of the PR across all of these energy scales, a symmetric logarithmic scale is used for the abscissa, where the portion of the axis near zero is scaled linearly, while the portions far from zero are scaled logarithmically. The size of the linear portion is denoted on the abscissa of the graph and in the captions.

\subsection{$\alpha = 0$}
\subsubsection{Isotropic Hopping}
Figure \ref{PRalphaZeroIsotropic} (upper panel) shows the PR of every eigenstate, for multiple disorder realizations, plotted against the energy of the eigenstate for $W=16.5$. The band at low energy is of width $16.5\tilde t$ and arises from the $(N^3-1)$-fold degenerate ground state of the ideal lattice. The points at the top of the spectrum arise from the $\ket{000_k}$ state of the ideal system. Figure \ref{PRalphaZeroIsotropic} (lower panel) shows the average PR as a function of energy for the band at low energies. There is clearly no scaling of $\langle\text{PR}\rangle$ with the system size. The fit to the $\langle\text{PR}\rangle$ of the high energy $\ket{000_k}$ state shows the scaling exponent to be 3. Identical behaviour has been observed for $W=1 \text{ and } 35$.
 
 Evidently, the band at low energy is composed of localized states, while the $\ket{000_k}$ state remains delocalized, even in the presence of strong disorder. Such behaviour is easily understood from the energy level structure of the ideal system (described in Section \ref{iso_analytics}). The extreme degeneracy of the ground state allows any small amount of disorder to strongly couple the states, leading to localization. The $\ket{000_k}$ state remains delocalized, however, as the shielding gap is significantly larger than the disorder strength, minimizing coupling. Given that the shielding gap diverges with the system size, the $\ket{000_k}$ state must remain delocalized in the thermodynamic limit for any finite disorder strength, while the low energy states remain localized. The same behaviour is observed for particles with isotropic uniform hopping on the \emph{d}-dimensional simplex \cite{Ossipov2013}.
 
 Given that cooperative shielding allows the system to be described by an effective short-range Hamiltonian \cite{Celardo2016}, one would expect to see the localization transition near $W=16.5$, the critical disorder strength of the Anderson transition for nearest-neighbour hopping in 3D lattices. Indeed, the dynamics of the system studied by Celardo \etal~\cite{Celardo2016} is governed by an effective nearest-neighbour hopping Hamiltonian. However, they explicitly include a nearest-neighbour hopping term in their Hamiltonian, in addition to the long-range term. As it would be inconsistent with our generalization of the dipolar hopping amplitude 
 to arbitrary $\alpha$, we do not include an additional nearest-neighbour hopping term in our Hamiltonian (\ref{model}). Thus, the ground state of our system when $\alpha =0$ is extremely degenerate (essentially flat-band) and localizes given any finite disorder strength. 
 
\begin{figure*}[ht]
	\begin{center}
		{\bf\large{~~~~~Anisotropic, $\bm{\alpha = 0$}}}\\
		\includegraphics[scale=0.4]{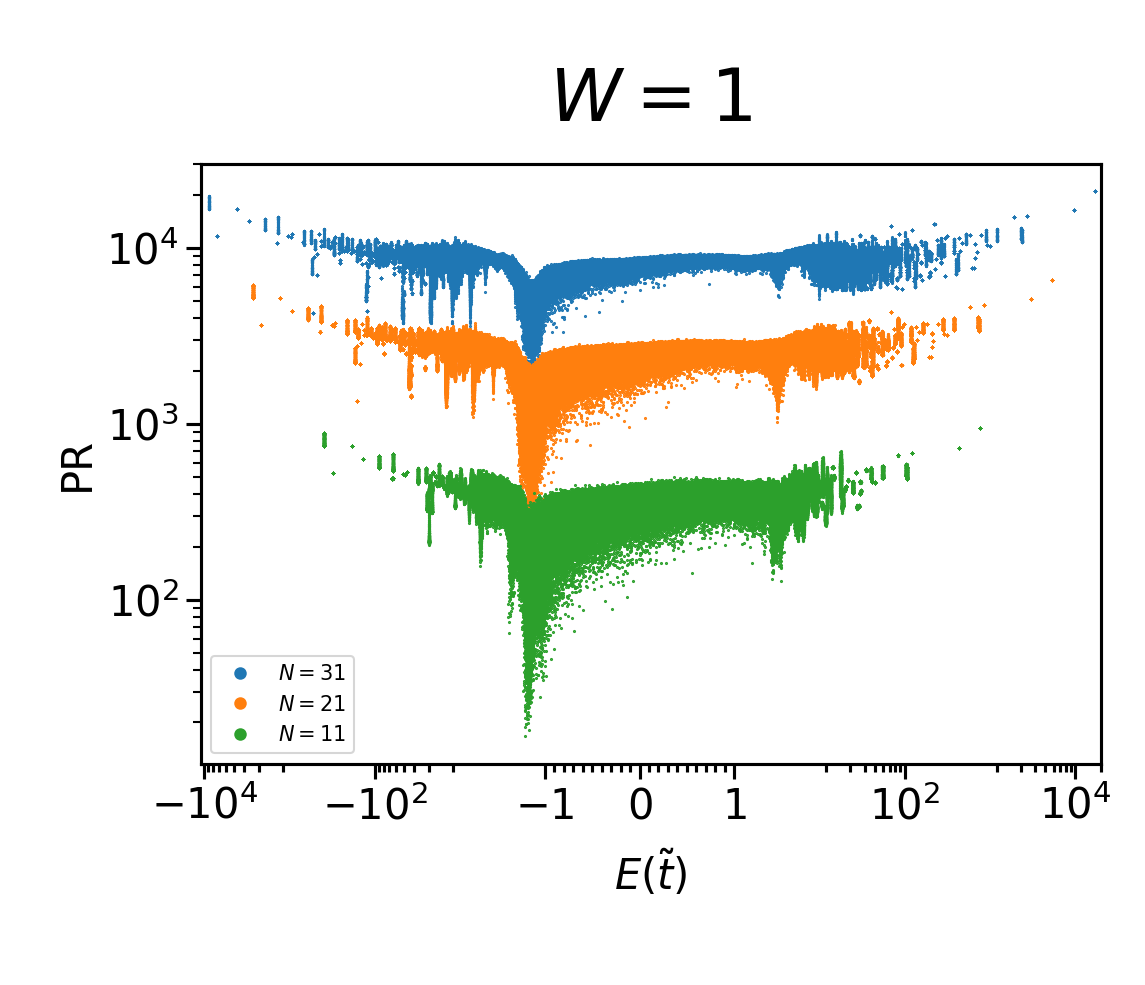}
		\includegraphics[scale=0.4]{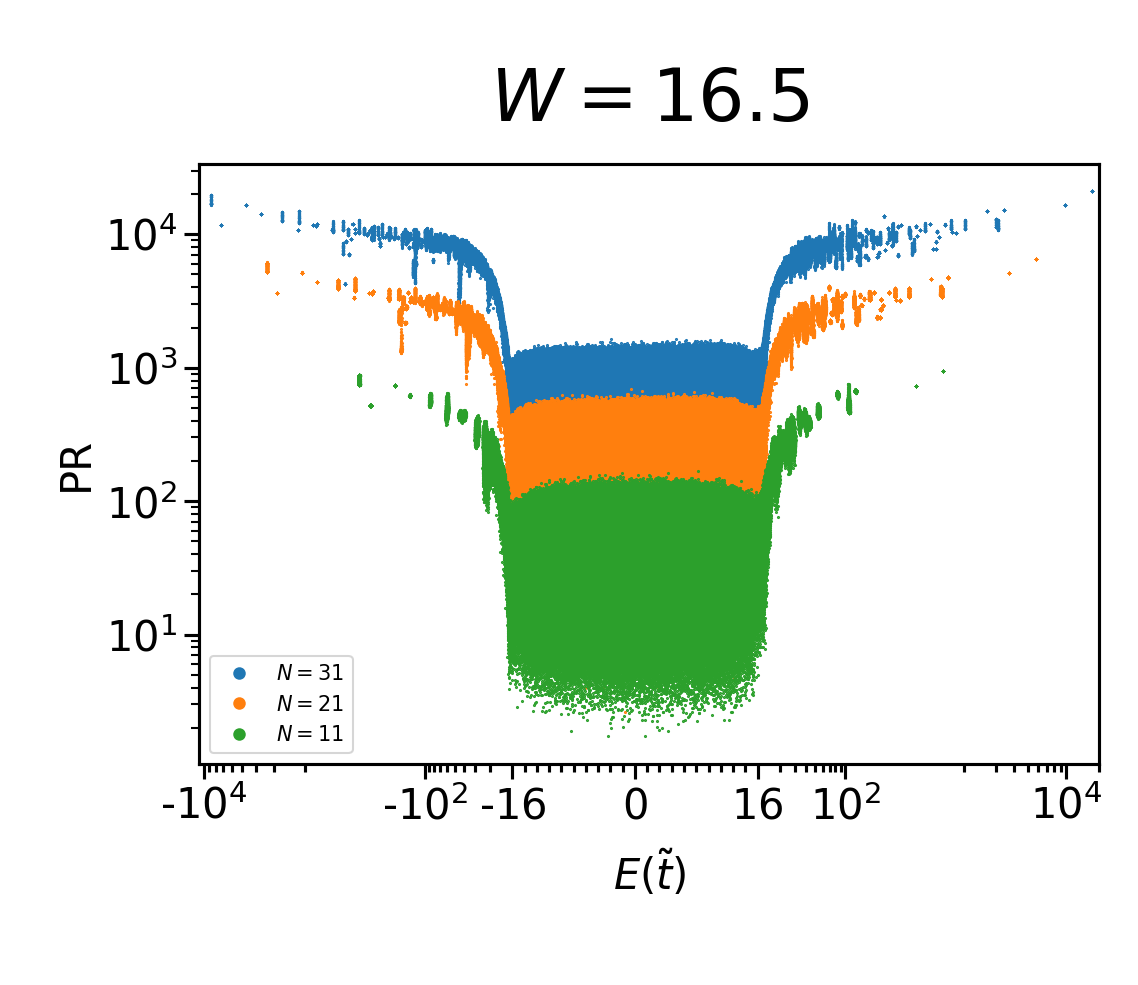} \\
		\includegraphics[scale=0.4]{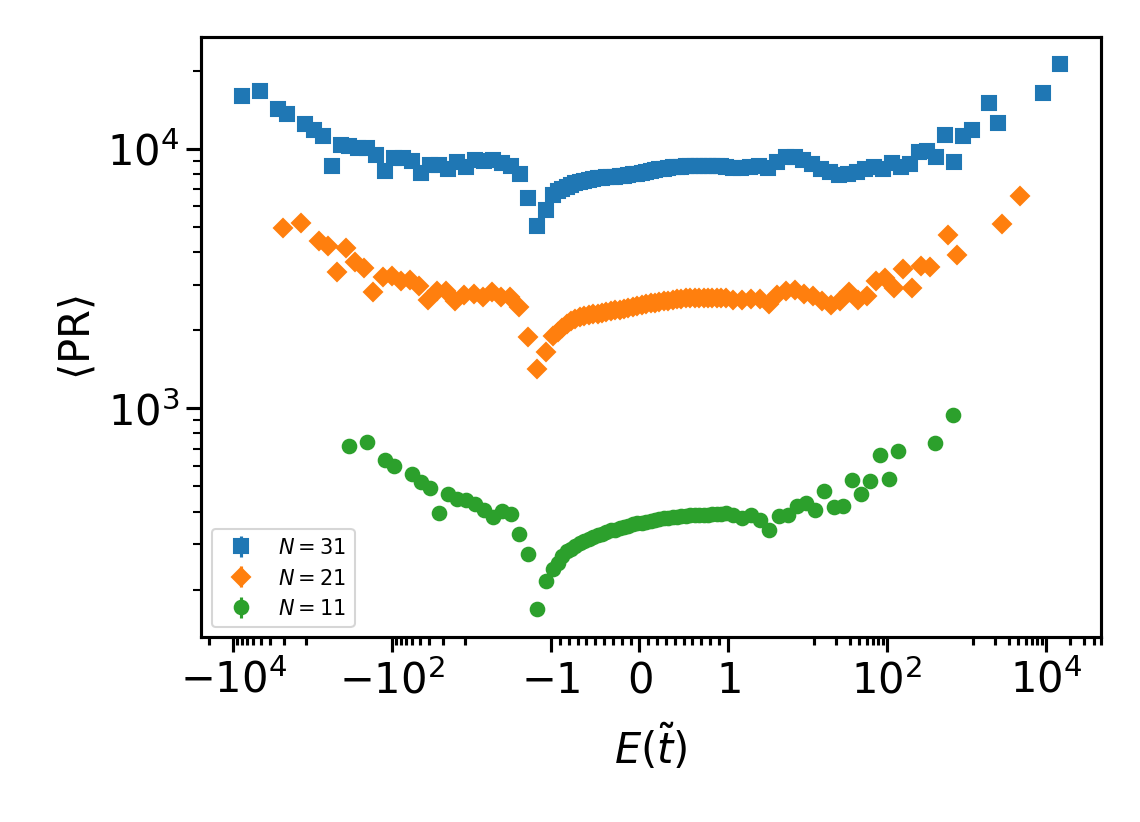} 
		\includegraphics[scale=0.4]{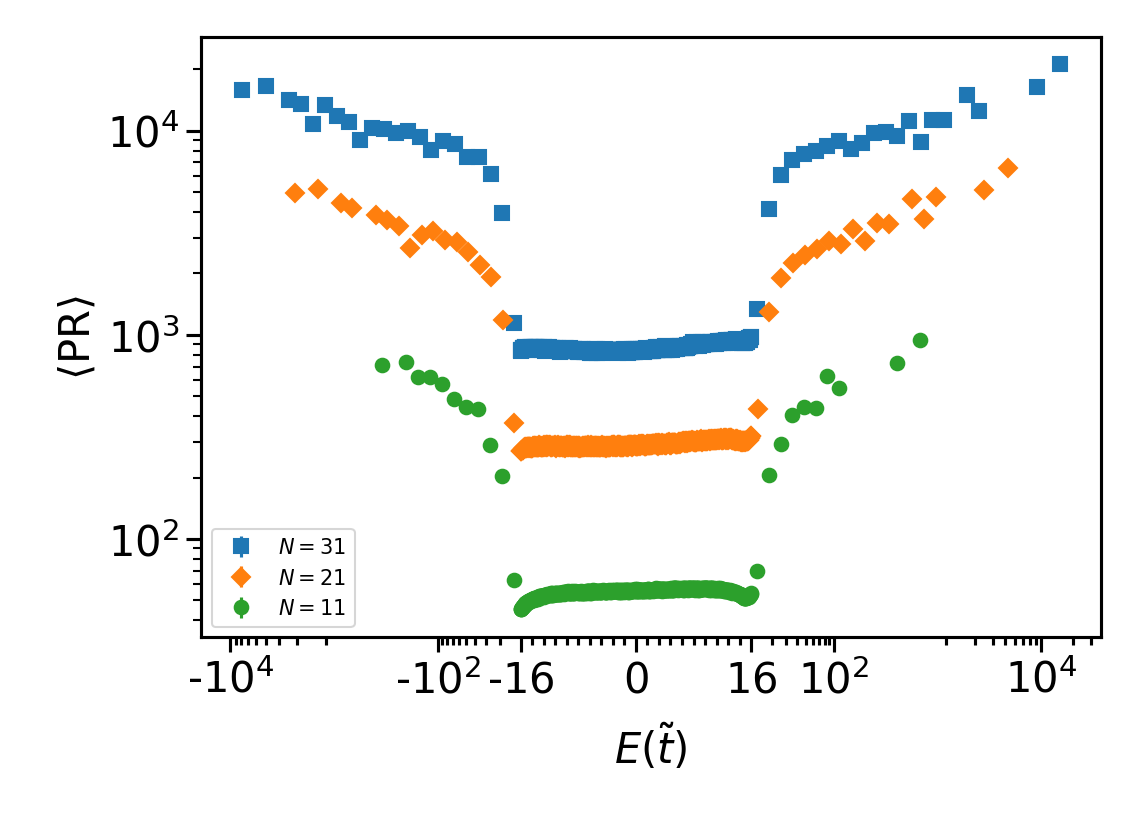} 
	\end{center}
	\caption{ Participation ratio (PR) as a function of energy for the eigenstate of a 3D system with anisotropic hopping; $W = 1 \text{ and } 16.5$; $p=1$; $\alpha=0$; and $N = 11,~21,\text{ and } 31$ (green \textcolor{fig_green}{$\blacksquare$}, orange \textcolor{fig_orange}{$\blacksquare$}, and blue \textcolor{fig_blue}{$\blacksquare$}, respectively). The number of disorders included ranges from 50 to 2350, as required to obtain sufficiently small error bars. \emph{Top Panels}: Plot of the PR of each eigenstate vs its eigenvalue for several disorder realizations.  \symLogScale{1}.  \emph{Lower Panels}: PR averaged within energy bins and over disorder realizations for $W = 1 \text{ and }16.5$. \symLogScale{16}. Error bars for the average PR are 95\% confidence intervals and smaller than the marker size where not seen.
	}
	\label{PRalphaZeroAnisotropic}
\end{figure*}
 
\subsubsection{Anisotropic Hopping}
The upper panels of Figure \ref{PRalphaZeroAnisotropic} show the PR vs. energy for $W = 1 \text{ and } 16.5$ (left and right, respectively), while the lower panels show the averaged PR at the same disorder strengths. One can clearly see scaling with the system size for both disorder strengths. Given that the bandwidth of the system is a function of the system size, the scaling behaviour cannot be quantitatively determined for all band energies. Clearly acceptable choices of energies for comparison include the top and bottom states of the band and the states near the centre. The wing regions, however, do not overlap well enough for a quantitative analysis. Fits to the scaling equation reveal that the top and bottom states at both disorders scale as $N^3$. The states in the energy region $[-1,1]$ for the system with $W=1$ also scale as $N^3$. Interestingly, the states in the centre portion of the system with $W=16.5$ have energy-dependent scaling exponents that vary from 2.6 to 2.8. We also examine the system $W=35$, but find it qualitatively identical to the $W=16.5$ case: the scaling exponents of the top and bottom states are 3, while those of the states near the band centre vary from 2.4 to 2.7. 

We note that the errors on the scaling exponents can be as large as 0.2. We obtain this upper bound on the error from one fit to the scaling equation that resulted in a value of 3.2, while $\beta$ cannot be larger than 3.  Given that we only have three system sizes that span a relatively small range of sizes, our scaling exponents  have low, but not negligible, precision.

Qualitatively, an examination of PR/$N^3$ (not shown) shows that the wings, for both disorder strengths, scale as $N^3$. The central states also scale as $N^3$ for $W=1$, while they scale with $0<\beta<3$ for the central states at $W=16.5$. These qualitative results support the accuracy of the quantitative results discussed above, despite their low precision.
 
 The above results show that the states in the wings are delocalized, even for strong disorder. Furthermore, the states near the centre of the band become extended non-ergodic for sufficient disorder strength. 
We also see that there are no localized states in this system, in great contrast to the isotropic case.

\begin{figure}[ht]
	\begin{center}
		{\bf\large{~~~~~$\bm{\alpha = 1$}}}\\
	\end{center}
	\begin{flushright}
		\includegraphics[scale=0.25]{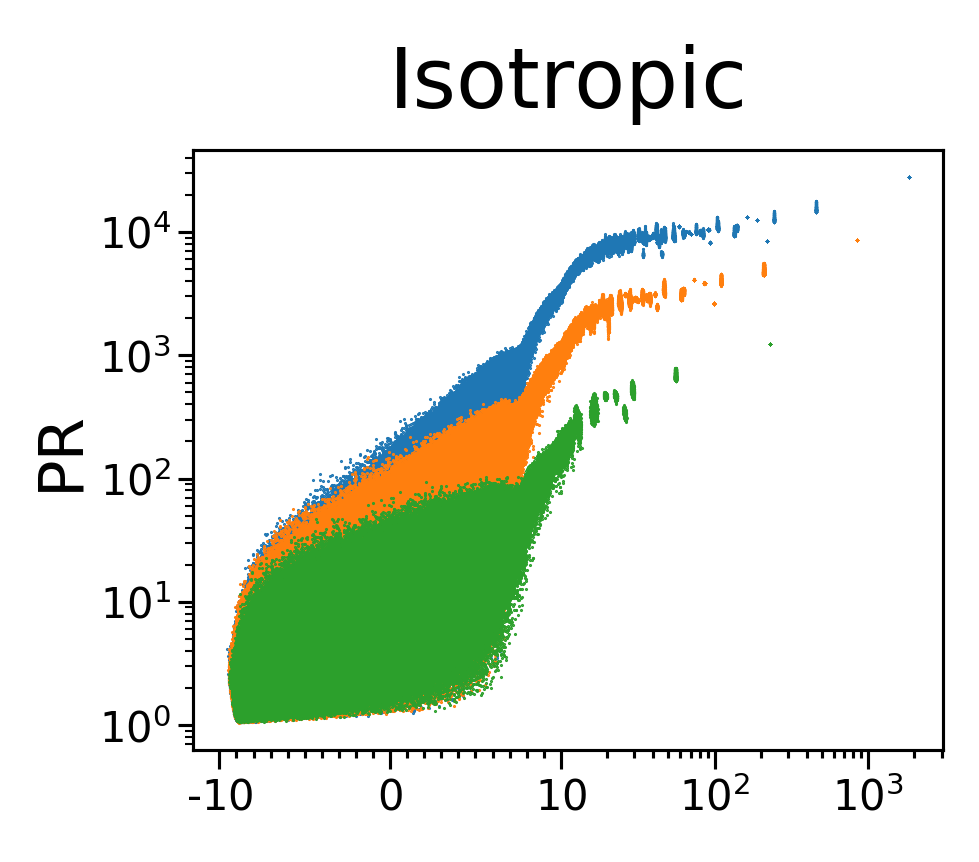}\phantom{f}
		\includegraphics[scale=0.25]{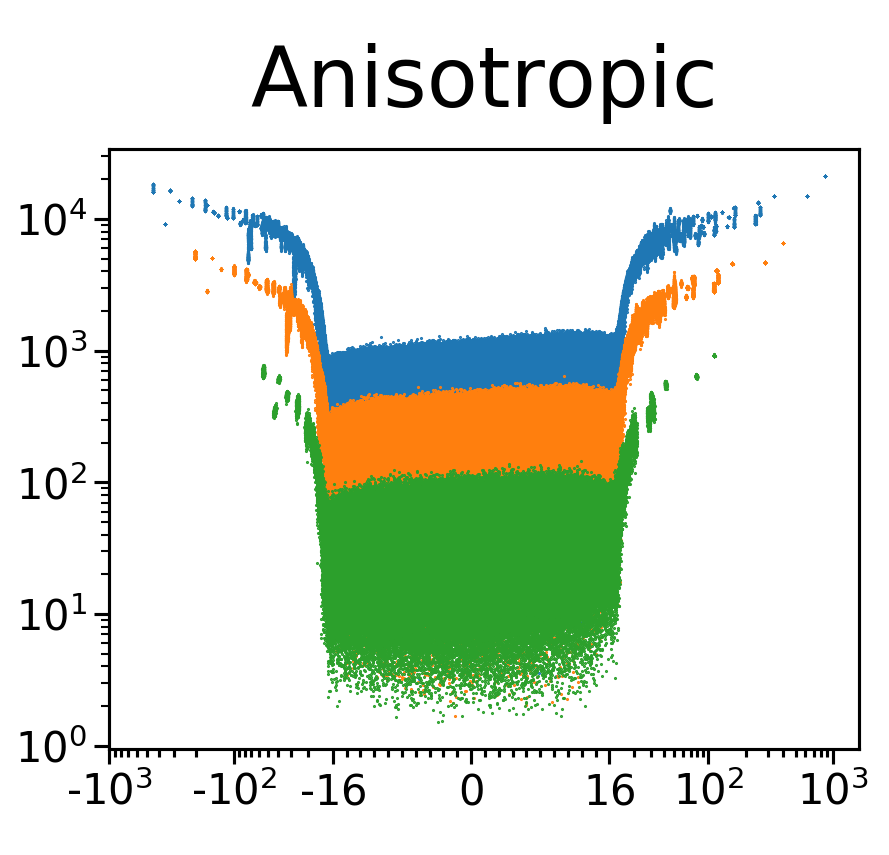} \\
		\includegraphics[scale=0.25]{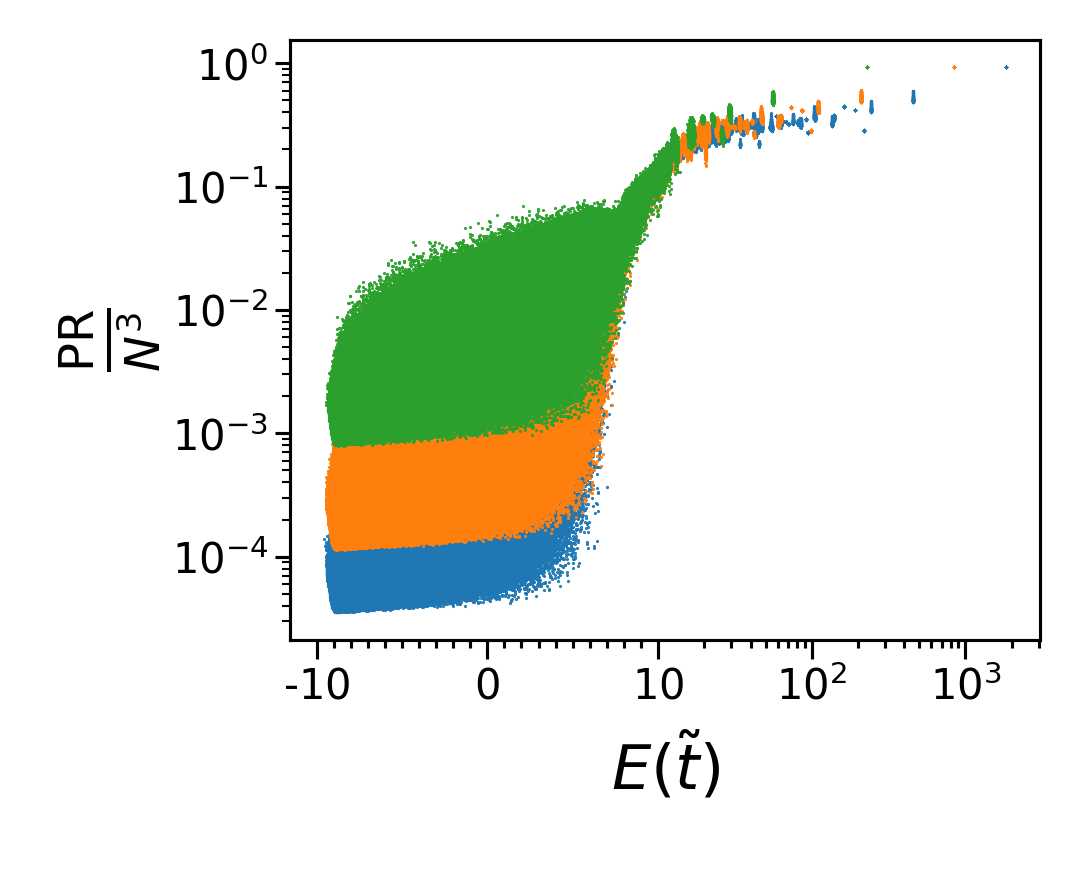} 
		\includegraphics[scale=0.25]{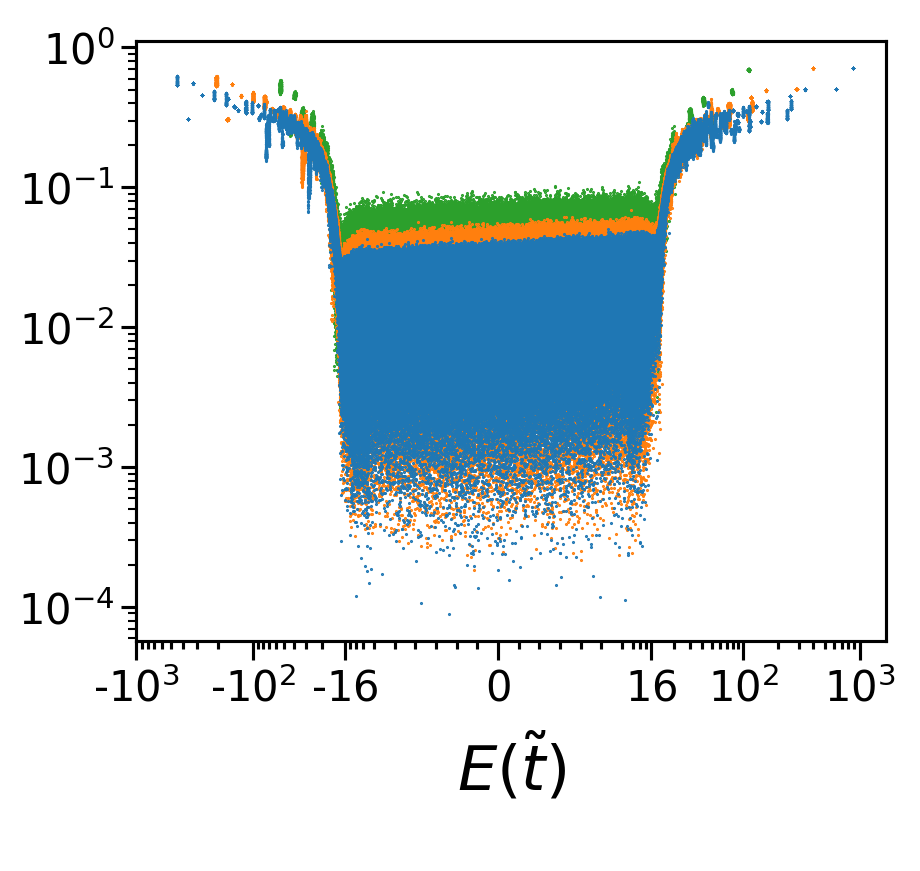} 
	\end{flushright}
	\caption{\emph{Top Panels}: Plot of the Participation ratio (PR) of each eigenstate vs its eigenvalue for several disorder realizations. \emph{Lower Panels}: Plot of the PR/$N^3$ of each eigenstate vs its eigenvalue for several disorder realizations. Plots are for a 3D system with isotropic and anisotropic hopping; $W = 16.5$; $p=1$; $\alpha=1$; and $N = 11,~21,\text{ and } 31$ (green \textcolor{fig_green}{$\blacksquare$}, orange \textcolor{fig_orange}{$\blacksquare$}, and blue \textcolor{fig_blue}{$\blacksquare$}, respectively). The number of disorders included ranges from 100 to 350, as required to obtain well-formed distributions. \symLogScale{10} or within $\pm16\tilde t$.
	}
	\label{PRalphaOne}
\end{figure}

\subsection{$\alpha = 1,3$}

We examine the scaling behaviour of the PR for isotropic and anisotropic hopping with $\alpha = 1$ and $\alpha =3$ by examining the qualitative scaling behaviour of PR and PR/$N^3$. We do not examine the quantitative behaviour as the limited system sizes accessible causes fits to the scaling equation to be even less reliable than when $\alpha=0$.

Figure \ref{PRalphaOne} demonstrates that, for $\alpha=1$ and $W=16.5$, there is a single wing for the isotropic case, while there are two for the anisotropic case. These wings contain PR values that increase with the system size, while the PR/$N^3$ values are constant: the values scale as $N^3$ and the states are delocalized. 

In the isotropic case near zero energy, however, the behaviour of the PR is not as clear. A portion of the PR values in this energy range does scale with the system size, but the distribution asymmetrically increases in width instead of shifting vertically. This obfuscates the interpretation. However, given that the PR/$N^3$ values all reduce in magnitude as the system size increases, it can be concluded that none of these central states are delocalized. Whether the states are extended non-ergodic or localized is not clear, however.

In the anisotropic case with $\alpha=1$, the states near zero energy are clearly extended non-ergodic. This is seen as both the PR and the PR/$N^3$ distributions shift vertically as the system size increases, in the appropriate directions.

Additionally, note that the rate of scaling with the system size near zero energy is different between the isotropic and anisotropic cases. The isotropic case has PR distributions that scale slowly, while the PR/$N^3$ distributions scale quickly. The reverse is true for the anisotropic case. We can thus conclude that the isotropic hopping scaling exponent is smaller than the anisotropic hopping scaling exponent for the states near zero energy.

\begin{figure}[ht]
	\begin{center}
		{\bf\large{~~~~~~~$\bm{\alpha = 3$}}}\\
	\end{center}
	\begin{flushright}
		\includegraphics[scale=0.24]{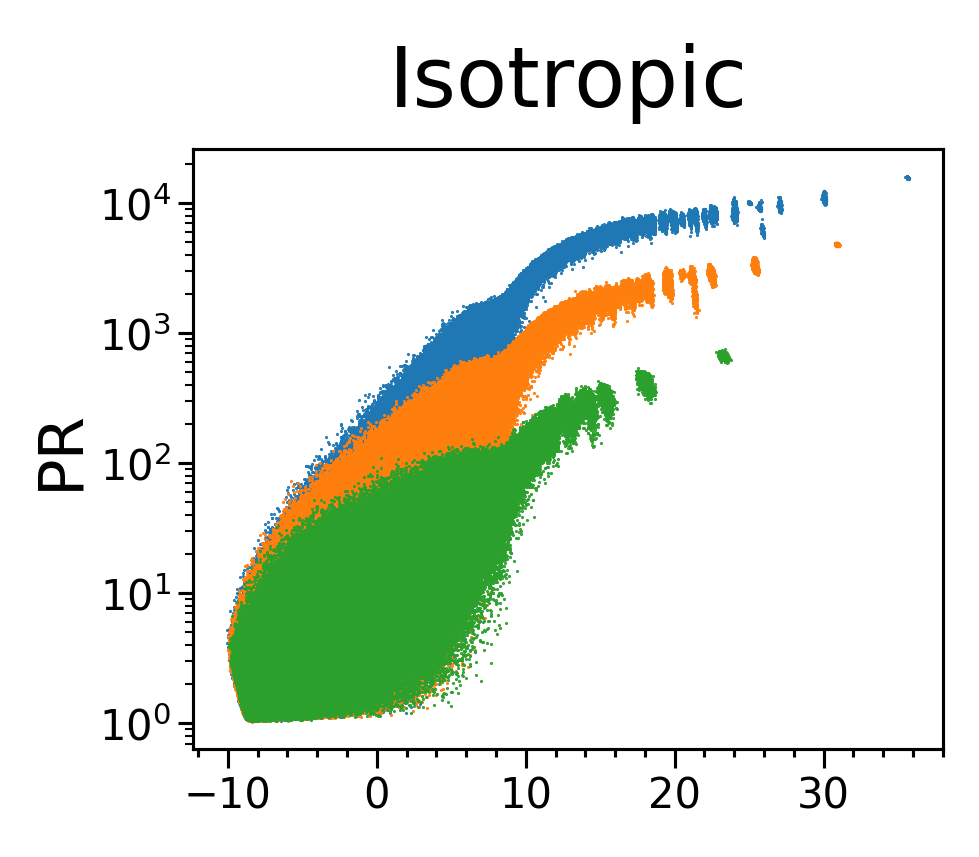}\phantom{f}
		\includegraphics[scale=0.24]{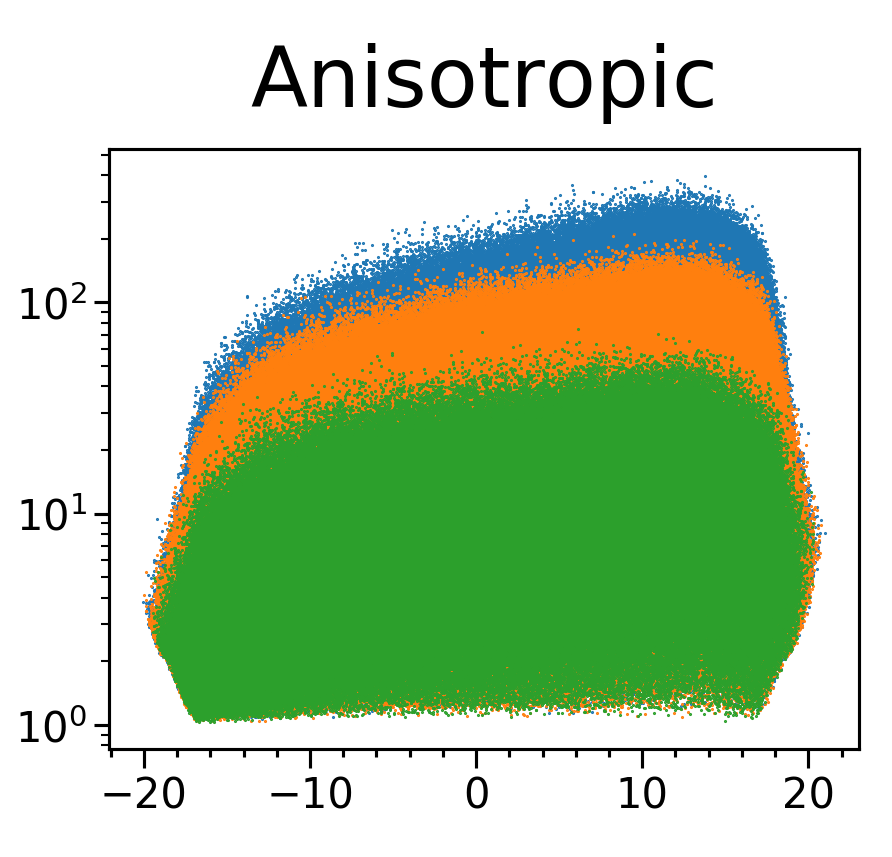} \\
		\includegraphics[scale=0.24]{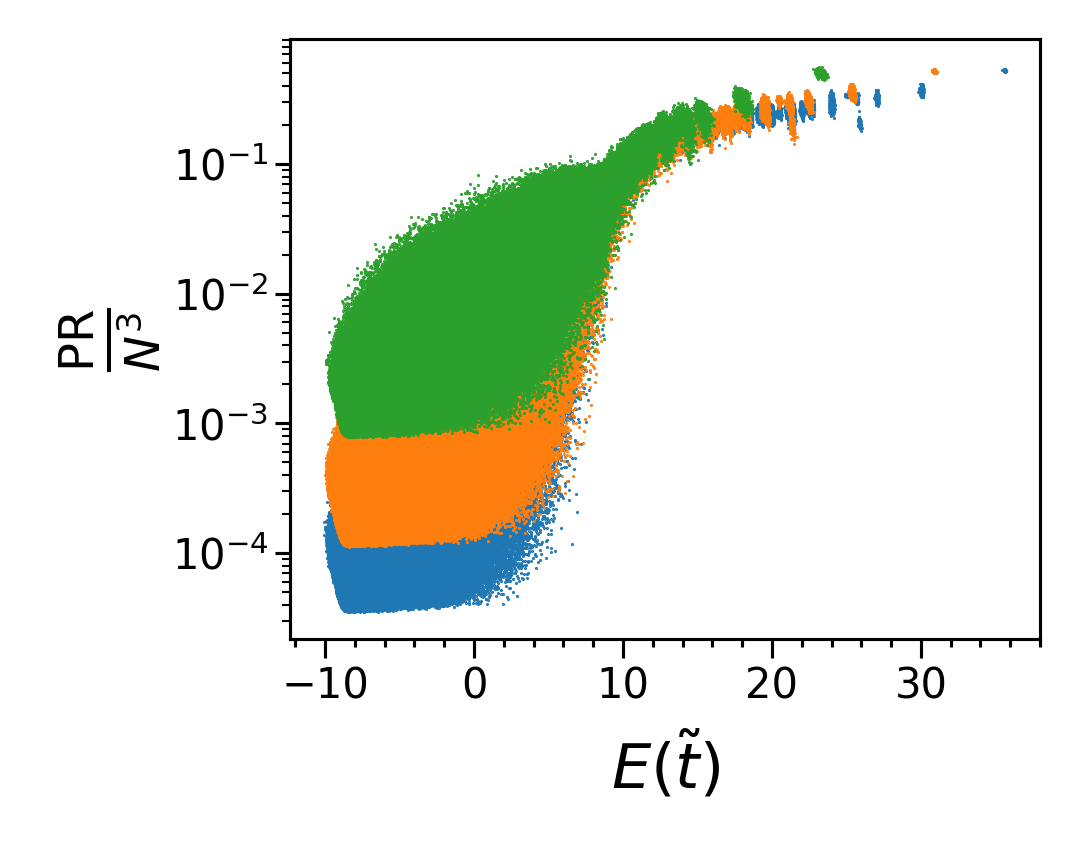} 
		\includegraphics[scale=0.24]{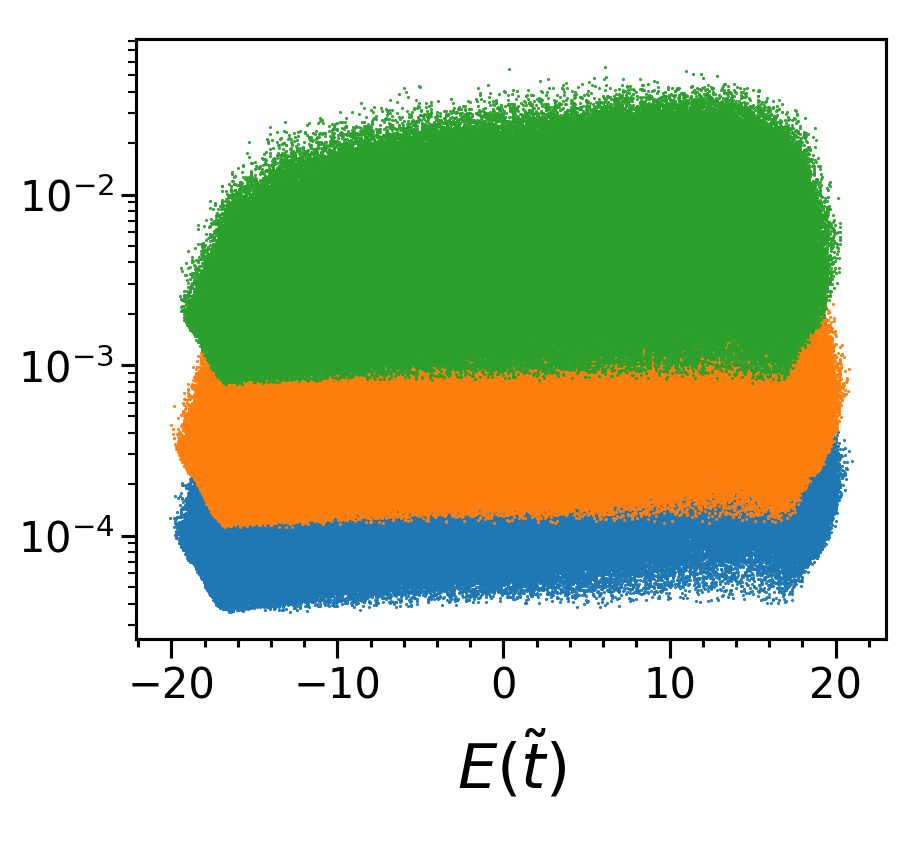} 
	\end{flushright}
	\caption{\emph{Top Panels}: Plot of the Participation ratio (PR) of each eigenstate vs its eigenvalue for several disorder realizations. \emph{Lower Panels}: Plot of the PR/$N^3$ of each eigenstate vs its eigenvalue for several disorder realizations. Plots are for a 3D system with isotropic and anisotropic hopping; $W = 16.5$; $p=1$; $\alpha=3$; and $N = 11,~21,\text{ and } 31$ (green \textcolor{fig_green}{$\blacksquare$}, orange \textcolor{fig_orange}{$\blacksquare$}, and blue \textcolor{fig_blue}{$\blacksquare$}, respectively). The number of disorders included ranges from 99 to 350, as required to obtain well-formed distributions. Note the linear scaling of the abscissa.
	}
	\label{PRalphaThree}
\end{figure}

Figure \ref{PRalphaThree} demonstrates that, for $\alpha=3$ and $W=16.5$, there is a single wing for the isotropic case, while there are none for the anisotropic case. Also note that the bandwidths are much smaller, allowing for a linear scaling of the x-axis. The behaviour of the isotropic case is similar to that seen for $\alpha =1$ and points to the same conclusion: the wing contains delocalized states, while the energy region near zero contains states that are not delocalized.  

At this level of disorder, the spectrum of the anisotropic case is dominated by the disorder, producing a rather uniform distribution without wings. Interestingly, the scaling behaviour is similar to the near-zero-energy portion of the isotropic case: the PR distributions widen asymmetrically but do not shift in location as the size increases, while the PR/$N^3$ distributions shift downward as the system size increases. There are thus no delocalized states, though there may be a combination of localized and extended non-ergodic states. This is not necessarily in contrast with the prior work of Levitov \cite{Levitov1989,Levitov1990,Levitov1990b} as the extended non-ergodic states still span the system, allowing weak transport.

The above analysis of the participation ratios for various values of $\alpha$ and for different isotropies shows that single-particle systems with long-range hopping display a rich landscape of states, from delocalized to extended non-ergodic to localized. We expect that the extended non-ergodic states we typically find near zero energy in this work are likely multi-fractal. To find and quantify the the multi-fractal dimensions and multi-fractal spectra \cite{DeLuca2014} of these states, however, will require more sophisticated numerical or analytical techniques  capable of accessing larger system sizes. There are algorithms that can reach system sizes of $N\approx300$ for targeted energy ranges \cite{Vasquez2008,Rodriguez2008}, but it is not clear whether even these would be enough to properly study the scaling behaviour of the eigenstates. 



\section{Energy Level Statistics}
Given the ambiguity observed in the scaling behaviour of the participation ratios for the physically important case of $\alpha =3$, we now examine the system for both isotropic and anisotropic hopping and open boundary conditions from an energy level statistics point of view.


We use the mean level spacing ratio, defined by Oganesyan and Huse \cite{Oganesyan2007} as 
\begin{align}
\langle r \rangle = \langle \min(r'_n,\frac{1}{r'_n})\rangle
\end{align}
where
\begin{align*}
r'_n = \frac{E_{n+1}-E_{n}}{E_{n}-E_{n-1}},
\end{align*}
$E_n$ is the $n$th eigenvalue, sorted in order of increasing magnitude and $\langle \cdot \rangle$ denotes the average over all eigenvalues and over many instances of disorder. For a system whose energy level spacing distribution is Poisonnian, $\langle r \rangle = 2\ln2 -1 \approx 0.38629$. If, however, the system belongs to the Gaussian Orthogonal Ensemble (GOE), $\langle r \rangle = 4 -2\sqrt{3} \approx 0.53590$ \cite{Atas2013}. As discussed in Refs. \cite{Oganesyan2007,Shklovskii1993},  Poissonian level statistics are characteristic of localization and  GOE statistics are characteristic of diffusion.

\begin{figure}[ht]
	\begin{center}
		{\bf\large{~~~~~~Full Spectrum}}\\
	\end{center}
	\begin{center}	
		\includegraphics[scale=0.4]{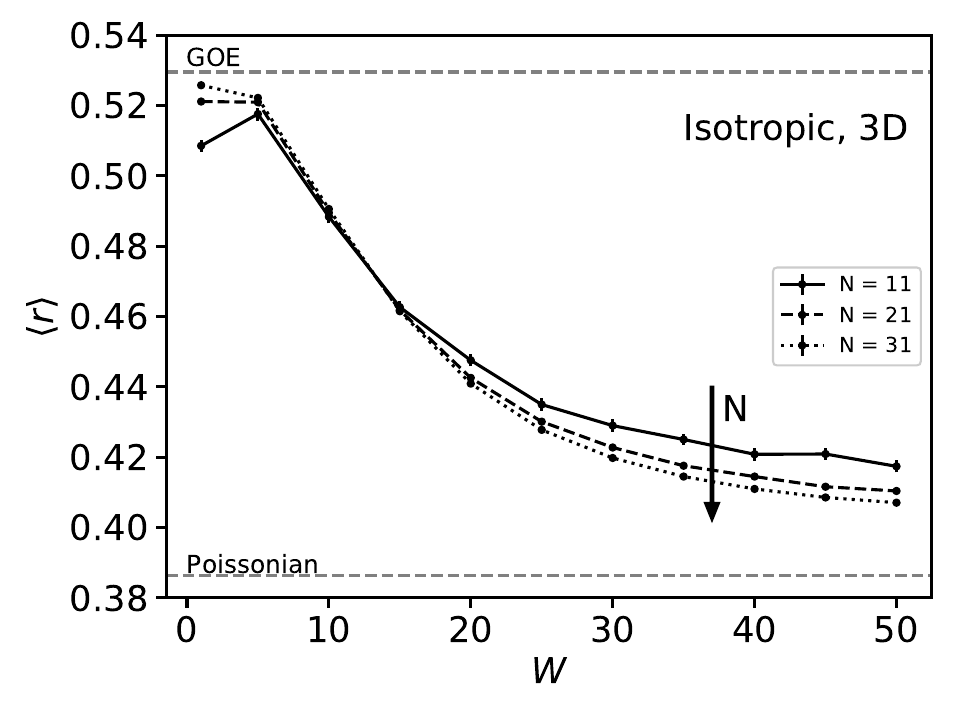} \\
		\includegraphics[scale=0.4]{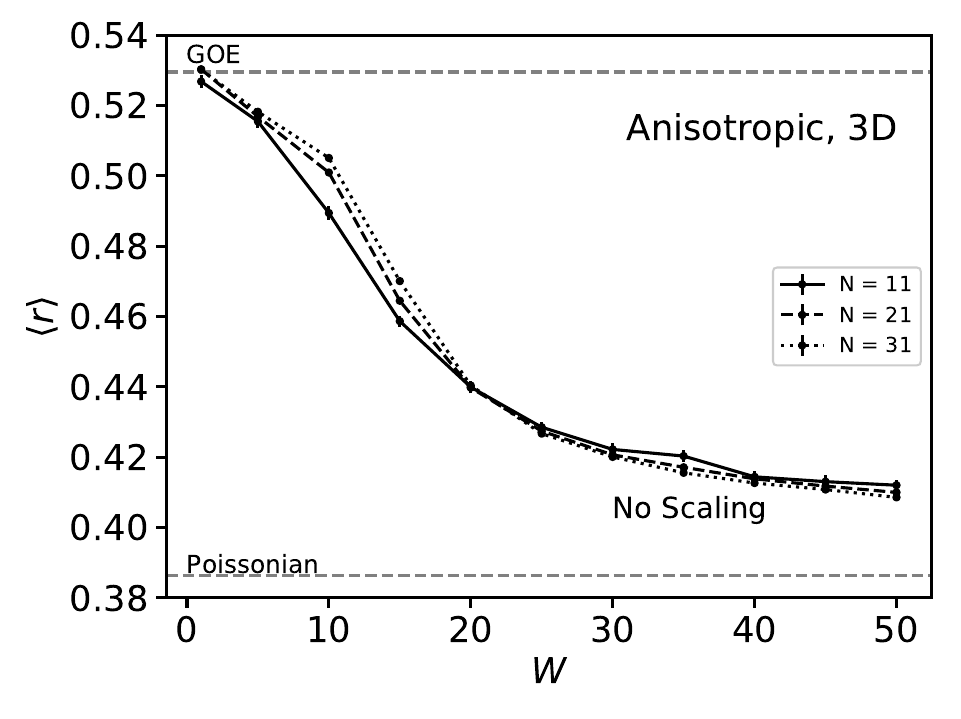} \\
	\end{center}
	\caption{
		Mean energy level spacing ratio $\langle r \rangle$ as a function of disorder strength $W = \frac{\omega}{t_{max}}$ for the isotropic and anisotropic hopping Hamiltonians with open boundary conditions. The hopping range exponent is $\alpha = 3$ and the filling fraction is $p=1$ in both cases. $N$ denotes the lattice side-length ($N^3$ sites). 
		The horizontal dashed lines denote the values of $\langle r \rangle$ for the Poisson distribution $({\sim} 0.38629)$ and the GOE $({\sim} 0.53590)$. The error bars are 95\% confidence intervals based on 96 disorders and are smaller than the marker size where not seen.
	}
	\label{1DhusParamPlot}
\end{figure}

Figure \ref{1DhusParamPlot} presents $\langle r \rangle$ vs $W$ for the isotropic and dipolar Hamiltonians with $\alpha = 3$ and open boundary conditions. It can be seen in both cases that $\langle r \rangle$ drops as $W$ increases, though not completely to the Poissonian value. However, in the isotropic case, $\langle r \rangle$ drops with the increasing lattice size towards the Poissonian limit. As illustrated in Figure \ref{1DhusParamPlot} (top panel), we observe a significant change with the system size. 
In contrast, in the anisotropic case, the curves obtained for different lattice sizes collapse into a single limit, even in the case of extremely strong disorder. There is thus no indication that the Poissonian limit will be attained.

It is noteworthy that the scaling behaviour in the isotropic case begins at $W\approx 15$, which is the expected location, given cooperative shielding. Cooperative shielding causes the Hamiltonian to be effectively short-ranged, such as the tight-binding model. For the tight-binding model, the diffusion-to-localization transition occurs at $W \approx 16.5$ \cite{root_skinner}.
At disorder strengths between 5 and 15, the curves in the isotropic case collapse onto one another, preventing the possibility of scaling to the infinite size limit. At $W = 1$, it is clear that the system is approaching the GOE value as the system size increases. 

\begin{figure}[ht]
	\begin{center}
		{\bf\large{~~~~~Near Zero Energy}}\\
	\end{center}
	\begin{center}
		\includegraphics[scale=0.4]{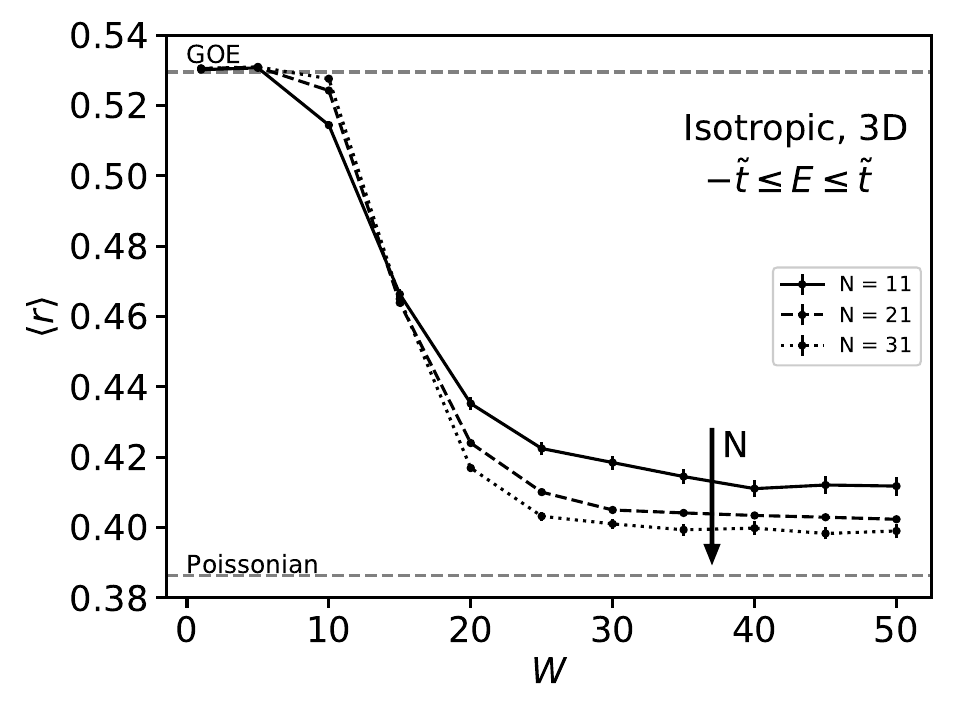} \\
	\end{center}
	\caption{
		Same as Figure \ref{1DhusParamPlot} (top panel), but with the mean energy level spacing ratio $\langle r \rangle$ including only the eigenvalues in the energy range $-\tilde t \leq E \leq \tilde t $.
		There are 1071, 1071, and 96 disorders for $N^3=11^3$, $21^3$, and $31^3$, respectively. 
	}
	\label{1DhusParamPlotTargetRange}
\end{figure}

Given that the scaling behaviour of the PRs reveals two clear regimes in the isotropic case, one where the states are delocalized and one where they are localized or extended non-ergodic (see Figure \ref{PRalphaThree}), it is useful to focus on the region near zero energy. Figure \ref{1DhusParamPlotTargetRange} demonstrates that the scaling behaviour observed previously is enhanced when the energy level spacing ratio is averaged only over the energy region $-\tilde t \leq E \leq \tilde t $.

These results provide numerical evidence for localization in the infinite size limit for the case with isotropic long-range hopping and $\alpha=3$. This, combined with the likely presence of cooperative shielding and the observed scaling behaviour of the participation ratios, strongly suggests that some, to many, of the states at low energy are localized.

\begin{figure}[ht]
	\begin{center}
		\includegraphics[scale=0.4]{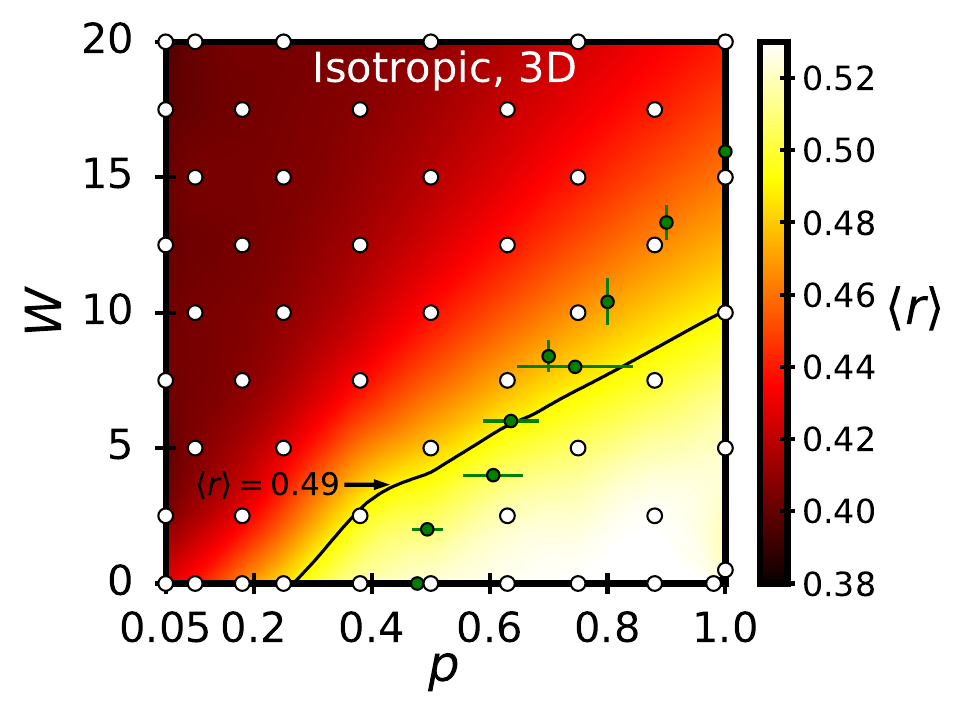} \\
		\includegraphics[scale=0.4]{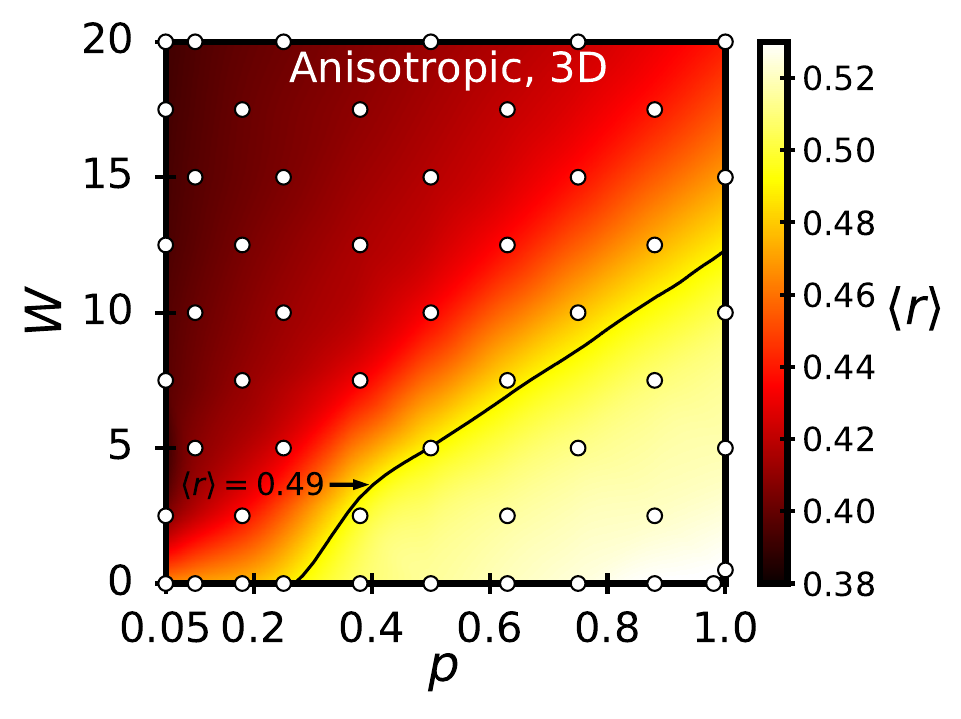} \\
	\end{center}
	\caption{
		Mean energy level spacing ratio $\langle r \rangle$ as a function of disorder strength $W = \frac{\omega}{t_{max}}$ and filling fraction $p$ for the isotropic and anisotropic hopping Hamiltonians with open boundary conditions. The hopping range exponent is $\alpha = 3$ and the lattice size is $N^3=31^3$ in both cases. The colour varies from the value of $\langle r \rangle$ for the Poisson distribution $({\sim} 0.38629)$ to that for the GOE $({\sim} 0.53590)$.  The white circles denote the values obtained via exact diagonalization of the Hamiltonian; piecewise cubic interpolation is used for the rest of the plot. The black lines indicate where $\langle r \rangle=0.49$. The results are averaged over 96 realizations of disorder, giving 95\% confidence intervals that are at most $\pm 0.002$. The green data points mark the location of the Anderson transition for the 3D tight-binding model in the infinite size limit, as determined by Root and Skinner \cite{root_skinner}.
	}
	\label{2DhuseParamPlot}
\end{figure}

To connect with realistic experimental conditions, we also examine the effect of a diluted lattice. This is relevant for experiments with polar molecules in an optical lattice \cite{jun-ye-2,jun-ye-nature} or for amorphous molecular solids \cite{Mladenovic2015}. A diagram of the mean level spacing ratio, $\langle r \rangle$, as a function of both $p$ and $W$ for both isotropic and dipolar hopping with $\alpha = 3$ and open boundary conditions is presented in Figure \ref{2DhuseParamPlot}.

While the two plots are qualitatively similar, $\langle r \rangle$ falls off more quickly with the disorder strength and filling fraction in the isotropic case. This can be seen by comparing the areas of the two diagrams where $\langle r \rangle > 0.49$. The value of $\langle r \rangle$ remains high along the $p$-axis, which is in agreement with the results of Deng \etal~\cite{LevyFlights}, who studied the case of dipolar hopping in a diluted lattice.

Figure \ref{2DhuseParamPlot} (bottom panel) shows that in experiments with polar molecules in optical lattices one should not expect to see localization of single rotational excitations due to the dilution alone, if it exists at all. Additional on-site disorder, such as from an optical speckle potential, is required. Exploring this phase diagram is within reach of current experiments. For example, Figure \ref{2DhuseParamPlot} (bottom panel) shows that for molecules on an optical lattice with $N \approx 30$ and a lattice population of 30 \% \cite{jun-ye-nature} (a typical filling fraction aimed at in current experiments), the region where $\langle r \rangle$ drops significantly can be explored by varying the optical speckle potential from below to above $W = 5$. Scaling behaviour of $\langle r \rangle$ can also be investigated as optical lattices can have $N$ up to 60 (a typical size of an optical lattice).

\section{Conclusion}

We have illustrated that the cooperative shielding discovered for 1D lattices with long-range hopping \cite{Santos2016,Celardo2016} also occurs in 3D lattice models with isotropic long-range hopping. This suggests the possibility of Anderson localization in 3D systems with long-range hopping (i.e.~$\alpha \leq 3$). Given the form of the results in three dimensions, the same is likely to be true of a 2D system with isotropic long-range hopping ($\alpha \leq 2$). We have also presented evidence indicating the lack of cooperative shielding in models with anisotropic dipolar hopping.

We have demonstrated that there are fundamental differences between disordered lattice systems with isotropic long-range hopping and those with anisotropic long-range hopping, particularly with regards to the impact of disorder. In addition to the difference in cooperative shielding, we have shown that the energy level structures of systems with isotropic hopping are qualitatively and quantitatively different from those of systems with dipolar anisotropic hopping (whether with periodic or open boundary conditions). 

We have used the scaling behaviour of the eigenstate participation ratios to demonstrate the \emph{presence} of localized states in the isotropic case with uniform hopping ($\alpha = 0$) and the \emph{absence} of localized states in the anisotropic case with uniform hopping ($\alpha = 0$). We have also demonstrated that energy level statistics support the presence of localized states in 3D systems with isotropic hopping that varies as $r^{-3}$, while they are inconclusive for systems with anisotropic dipolar hopping. 

We have shown that the localization properties (or lack thereof) found for systems with dipolar hopping \cite{Levitov1989,Levitov1990,Levitov1990b,long-range-effects} cannot be assumed to apply to systems with isotropic hopping, in accordance with the suggestion of Burin and Maksimov \cite{Burin1989}. More generally, we have demonstrated that the presence or absence of delocalized, extended non-ergodic, or localized states depends on both the hopping exponent $\alpha$ and the isotropy of the hopping amplitudes. 

Future studies that can access significantly larger system sizes are required to fully characterize the shape of  the localized states, to determine whether the extended non-ergodic states are truly multi-fractal, and to determine how the multi-fractal dimensions and spectra change with hopping range and isotropy.
Furthermore, this work opens up the question of how different types of isotropy, such as quadrupolar-like, can impact localization behaviour.

\section*{Acknowledgments}
We acknowledge useful discussions with Andreas Buchleitner and Alexander L. Burin. This work is supported by NSERC of Canada.

\end{document}